\documentclass[12pt]{article}

\usepackage[english]{babel}
\usepackage[letterpaper,top=2cm,bottom=2cm,left=1.8cm,right=1.8cm,marginparwidth=1.5cm]{geometry}
\usepackage{amsmath}
\usepackage{graphicx}
\usepackage[colorlinks=true, allcolors=black]{hyperref}
\usepackage{bm}
\usepackage{float}
\usepackage{hyperref}
\usepackage{bbold}
\usepackage{verbatim}
\usepackage[table]{xcolor}
\definecolor{highlight}{RGB}{135,206,250} 
\usepackage{subfigure}
\usepackage{caption}
\usepackage{multirow}
\usepackage{booktabs}
\usepackage{natbib}
\usepackage{tikz}    
\newcommand{\hide}[1]{}
\usetikzlibrary{shapes.geometric, arrows, shadows, fadings}
\tikzstyle{startstop} = [rectangle, rounded corners, minimum width=3.5cm, minimum height=1.2cm, text centered, draw=black, fill=gray!20]
\tikzstyle{arrow} = [thick,->,>=stealth]

\title{A Bayesian multivariate model with temporal dependence \\ on random partition of areal data}

\author{Jessica Pavani \& Fernando Andr{\'e}s Quintana \\
	  \small Departamento de Estad{\'i}stica, Pontificia Universidad Cat{\'o}lica de Chile
}

\date{}

\begin{document}

\maketitle

\begin{abstract}
More than half of the world's population is exposed to the risk of mosquito-borne diseases, which leads to millions of cases and hundreds of thousands of deaths every year. Analyzing this type of data is often complex and poses several interesting challenges, mainly due to the vast geographic area, the peculiar temporal behavior, and the potential correlation between infections. Motivation stems from the analysis of tropical diseases data, namely, the number of cases of two arboviruses, dengue and chikungunya, transmitted by the same mosquito, for all the 145 microregions in Southeast Brazil from 2018 to 2022. As a contribution to the literature on multivariate disease data, we develop a flexible Bayesian multivariate spatio-temporal model where temporal dependence is defined for areal clusters. The model features a prior distribution for the random partition of areal data that incorporates neighboring information, thus encouraging  maps with few contiguous clusters and discouraging clusters with disconnected areas. The model also incorporates an autoregressive structure and terms related to seasonal patterns into temporal components that are disease and cluster-specific. It also considers a multivariate directed acyclic graph autoregressive structure to accommodate spatial and inter-disease dependence, facilitating the interpretation of spatial correlation.  We explore properties of the model by way of simulation studies and show results that prove our proposal compares well to competing alternatives. Finally, we apply the model to the motivating dataset with a twofold goal: clustering areas where the temporal trend of certain diseases are similar, and exploring the potential existence of temporal and/or spatial correlation between two diseases transmitted by the same mosquito. 
\end{abstract}

\thispagestyle{empty}

\newpage
\setcounter{page}{1}

\section{Introduction} \label{s:intro}

In epidemiological investigation, it is crucial to study the link between geographical locations and/or temporal trends and the occurrence of diseases. Overall, disease incidence or mortality data are usually recorded as summary counts for contiguous geographical regions (e.g. census tracts, postcodes, districts, or counties) and collected over discrete time periods, usually epidemiological weeks. Responses are frequently accompanied by covariates describing information associated with regions and/or time periods. In this context, generalized linear mixed models, mainly Poisson regression, play a major role. Alternatively, it may be useful to apply a transformation in order to produce a continuous response, for instance, the popular Freeman-Tukey double sine transformation \citep{Freeman1950}, and thus, a Gaussian regression would be appropriate. However, the availability of increasingly greater data of higher quality has been bringing new challenges, which implies the emergence of new and more sophisticated statistical models and methods.

Models that describe the geographical distribution of diseases and their evolution over time are abundant in the spatio-temporal disease mapping literature. From the Bayesian perspective, two of the main contributions to this field were described by \cite{Besag1991} and \cite{Leroux2000} which are among the most popular specifications of conditionally autoregressive (CAR) models. This structure is frequently employed for representing spatial autocorrelation in areal data and it may be considered as a conditional description of a Gaussian Markov random field \citep{Rue2005}. The class of CAR models is large and many variations can be found \citep{Besag1974, Besag1975, Kunsch1987, Mardia1988, Besag1995}. Furthermore, many authors have concentrated on using CAR models and their variations \citep{Lawson2009, Banerjee2014, Blangiardo2015, Martinez-Beneito2019}. Even though CAR models are widely accepted in the statistical community, they may bring some concerns regarding identifiability and interpretation \citep{Goicoa2018}. As an alternative, \cite{Datta2019} proposed the directed acyclic graph auto-regressive model for areal datasets. In this approach, the adjacency matrix is built over a directed acyclic graph, which in turn is derived from the original undirected graph, i.e., an order is imposed on areas. Although it is quite recent, some extensions may already be found \citep{Gao2022, Gao2023, Aiello2023}.

The literature on modeling mosquito-borne diseases is also extensive. \cite{Aswi2018} conducted a systematic search to compare different spatial and spatio-temporal Bayesian methods applied to dengue. Similarly, there are many other studies \citep{Aswi2020, Lim2020, Rotejanaprasert2020, Ye2020}. However, although some arboviruses are transmitted by the same vector, literature focusing on multiple diseases is sparse. \cite{Carvalho2020} examined the connection between the Zika epidemic and past dengue outbreaks in the same region. In terms of clustering, \cite{Freitas2019} and \cite{Kazazian2020} explored simultaneous clustering patterns of dengue, Zika, and chikungunya epidemics. More recently, \cite{Schmidt2022} modeled simultaneous outbreaks in a specific area to assess the relationship between relative risk and environmental and socioeconomic factors. \cite{Pavani2022} and \cite{Pavani2023} conducted spatio-temporal and spatial studies to jointly model dengue and chikungunya aiming to identify shared patterns between diseases. Even so, in most of these works, the correlation between diseases was not taken into account. 

Despite the fact that spatial and spatio-temporal epidemiology have grown substantially in the past years, more flexible approaches have received comparatively less attention. \cite{Kottas2008} developed a hierarchical specification using spatial random effects modeled with a Dirichlet process prior. This approach allowed for the modeling of disease rate surfaces, considering both space and time, in a dynamic setting. To demonstrate the effectiveness of the model, Ohio lung cancer mortality data was used. \cite{Hossain2013}, developed a mixture model where spatial and temporal effects are introduced by using space-time covariate dependent kernel stick-breaking processes. Then, this modeling was applied to South Carolina low birth weight data. \cite{Cassese2019} proposed a species sampling model to monitor pneumonia and influenza mortality that permits the identification of disease outbreaks taking into account the spatio-temporal dependence. \cite{Wehrhahn2020} introduced a restricted Chinese restaurant process prior that constrains clusters to be made of adjacent areal units. Their model was then applied to oral cancer mortality in administrative districts in Germany. Lastly, to deal with correlated diseases, \cite{Aiello2023} adopted a class of multivariate areally referenced Dirichlet process models to accommodate both spatial and interdisease dependence. They used occurrences of four potentially interrelated types of cancers in California to illustrate their model.

Spatial clustering, which may involve areal clusters that change over time, is a promising field of research. Firstly, \cite{Hegarty2008} proposed a prior for the random partition of areal data based on the number of its neighbours. This construction encourages maps with few contiguous clusters and discourages clusters with disconnected areas. Subsequently, \cite{Teixeira2015} introduced random spanning trees into the random product partition model for areal clustering. Thus, they could reduce the search space of partitions and guarantee that only partitions that satisfy the geographical constraint were considered. Later, the same authors extended their strategy to the spatial-temporal context by building a tree that takes into account both spatial and temporal information \citep{Teixeira2019}.  Most recently, \cite{Cremaschi2023} introduced a spatial-temporal approach where the prior for the random partition of areal data combines the Dirichlet process and the specifications proposed by \cite{Hegarty2008}. To allow for changes over time, they incorporated a temporal change-point component.

Motivated by the lack of literature on flexible approaches for mosquito-borne disease, we develop an approach focused on this context. In particular, this manuscript seeks to explore the potential existence of correlation between two diseases transmitted by the same mosquito. To do so, we formulate a spatio-temporal model to identify and cluster areas where certain diseases behave similarly. Specifically, our contributions can be summarized as follows: (i) we develop a flexible multivariate spatio-temporal model where temporal dependence is defined for areal clusters induced by product partition models. In this way, we allow each cluster in each disease to have different temporal behaviors. (ii) The temporal component is versatile and allows the use of any order autoregressive models or even harmonic functions. Additionally, components that explain potential seasonal behaviors can also be included, which may be useful to model data from arboviruses. (iii) Our model admits spatial, temporal, and/or spatio-temporal covariates, which may impact each disease differently. (iv) We also incorporate spatial random effects that take into account the neighborhood structure of the areas. They are modeled using a class of directed acyclic graph autoregressive models that allows us to interpret the spatial autocorrelation of each disease as well as the association between them. Finally, (v) the proposed hierarchical model considers correlation between response, in both temporal and spatial levels. 

The remainder of this paper is structured as follows. Section~\ref{s:motivation} describes the context in which this study was motivated. In Section~\ref{s:model}, we introduce the model and discuss its main features. Section~\ref{s:simulations} is dedicated to simulation studies carried out to elicit insights into the performance of our proposed model using different metrics, in addition to comparing our results with competing alternatives. The findings by applying the model to mosquito-borne disease data are presented in Section~\ref{s:dataapp}. Finally, Section~\ref{s:conclusion} concludes the paper with a discussion. Computational details on a posterior simulation algorithm, extra results, and univariate analyzes are included in the Appendix.

\section{Mosquito-borne diseases and motivating dataset} \label{s:motivation}

Arboviruses is a general term used to describe infections caused by a virus spread by the bite of infected vectors, mainly mosquitoes. Among the main mosquito-borne diseases are: chikungunya, dengue, malaria, Zika, West Nile, and yellow fever. Although there are many arboviruses, some of their spread characteristics are similar. Indeed, some of the cited viruses are transmitted by the same mosquito species, whose breeding and development are usually influenced by climatic factors. Over the last decades, the frequency and magnitude of arboviruses outbreaks have increased dramatically, mostly in the underdeveloped tropical regions, exposing more than half of the world's population to risk. According to the World Health Organization, over 3 millions cases of mosquito-borne diseases were reported only in the region of the Americas in 2022, which represents an expressive increase compared to 2021, when almost 1.5 million cases were registered. The majority of arbovirus cases in the Americas, with dengue being the most common, are attributed to Brazil. After being eradicated, the virus reappeared in the country in the 1980s and has persisted ever since. Additionally, Chikungunya is another widespread disease transmitted by mosquitoes in the region. It was first reported in Brazil in 2014 in the states of Amap{\'a} and Bahia and currently both infections coexist in the country. These diseases are transmitted by the same mosquito species, {\it Aedes aegypti}, and people infected by them present similar symptoms. Hence, it is reasonable to think that these diseases are correlated or even there is a competitive suppression between viruses. Therefore, investigating this potential correlation is important and may help in public health decision-making.

In this study, we focus on dengue and chikungunya cases of the Brazilian Southeast region \cite[see][for further details]{Codeco2018}. This region consists of four states: Espirito Santo (ES), Minas Gerais (MG), Rio de Janeiro (RJ), and S{\~a}o Paulo (SP) totalling almost 90 million people. We used the second level of the Brazilian administrative division considered by the Brazilian Institute of Geography and Statistics (IBGE in portuguese), known as microregions. Thus, we have 145 areal units ($n = 145$, 8 from Espirito Santo, 70 from Minas Gerais, 14 from Rio de Janeiro, and 53 from S{\~a}o Paulo). The number of cases in each area was weekly counted from 2018 to 2022, leading to 261 epidemiological weeks ($T = 261$). For computational convenience, Freeman-Tukey double sine transformation was applied to the original data in order to produce continuous responses.

Analyzing these data is complex and poses several interesting challenges. Apart from the complexity due to the vast geographic area and the correlation between diseases, the temporal behavior of mosquito-borne diseases is peculiar. Transmitting mosquitoes need favorable climatic conditions, such as rainfall and warm temperatures, causing many cases in summer and few in winter. This seasonal behavior can be quite challenging since the proportion of zeroes increases significantly in certain periods (see Figure~\ref{f:empirical_dta} of Appendix~\ref{appB}). Furthermore, climatic conditions are not homogeneous across the country or even across each state, which suggests that areal clusters may change over time. Hence, environmental factors should be chosen as covariates in order to better capture the data behavior. After some exploratory analysis, we observed that considering seasonal indicators can be a good way to capture the temporal behavior of dengue and chikungunya data, since the season indicates climatic characteristics in a combined way. At this point, it could be important to highlight that Brazil is located in the Southern hemisphere, so summer occurs from December to March, followed by autumn from March to June, winter from June to September, and finally, spring from September to December. It is also well known in the literature that socio-demographic factors are relevant in the context of mosquito-borne diseases. Thus, we incorporated the information coming from the Human Development Index (HDI) as a spatial covariate. HDI is a summary measure of average achievement of human development which  was obtained from the 2010 Demographic Census where each area unit has its own HDI value. 

\section{Model development} \label{s:model}

Before detailing our method, we introduce some general notation. Let $i = 1, \ldots, n$ denote the $n$ areal units at time $t$ for $t = 1, \ldots, T$, and disease $d = 1, \ldots, D$. Furthermore, let ${\bm \rho} = \{S_{1}, \ldots, S_{k}\}$ denote a partition of the $n$ areal units into $k$ clusters. An alternative notation is based on $n$ cluster labels, denoted by $c = \{c_{1}, \ldots, c_{n} \}$, where $c_{i} = j$ implies that $i \in S_{j}$ with $j = 1, \ldots, k$. Finally, any quantity with a “$\star$” superscript will be cluster-specific. Specifically, we use ${\bm \gamma}_{j}^{\star}$ to denote coefficients related to the temporal structure of cluster $j$ and $\sigma_{j}^{2 \star}$ to denote the variance of cluster $j$.

\subsection{Likelihood} \label{ss:modelspec}

Let $y_{itd}$ be the continuous outcome for area $i$ and disease $d$ at time $t$,
which we model as:
\begin{equation}  
(y_{itd} \mid X_{itd}, {\bm \beta}_{d}, Z_{itd}, {\bm \gamma}_{c_{i}d}^{\star}, \phi_{id}, \sigma_{c_{i}d}^{2 \star}) \overset{iid}{\sim} \text{N}(X_{itd}^{\top}{\bm \beta}_{d} + Z_{itd}^{\top}{\bm \gamma}_{c_{i}d}^{\star} + \phi_{id}, \, \sigma_{c_{i}d}^{2 \star}),
\label{eq:model} \end{equation} 
where ${\bm \beta}_{d} = \{ \beta_{1d}, \ldots, \beta_{pd} \}$ are regression coefficients related to a $p$-dimensional design vector, ${\bm X}_{itd}$, that considers spatio-temporal predictors for each disease. Similarly, ${\bm \gamma}_{jd}^{\star} = \{ \gamma_{j1d}^{\star}, \ldots, \gamma_{jqd}^{\star} \}$ are coefficients related to a $q$-dimensional temporal design vector, ${\bm Z}_{itd}$. As ${\bm \gamma}_{jd}^{\star}$ are cluster-specific parameters, all areas belonging to the cluster $j$ share the same parameter values for their respective response. These quantities are also disease-specific, which allow different temporal trends to each cluster and disease. Besides, $\phi_{id}$ represents spatial random effect of area $i$ and disease $d$. Note that $\beta_{d0}$ and $\gamma_{jd0}^{\star}$ are not included in the model, so that the spatial effect $\phi_{id}$ plays also a role of random intercept, which makes the model completely identified. Finally, $\sigma_{jd}^{2 \star}$ is the data variance of areas belonging to the cluster $j$ for disease $d$.

To perform full Bayesian analysis using the likelihood given above, we consider prior distributions for all parameters, which in the case of regression coefficients and variances are assumed to be:
\begin{equation}
   {\bm \beta} \sim \text{N}_{pD} ({\bm \mu}_{\beta}, {\bm \Sigma}_{\beta}), \quad (\sigma_{jd}^{2 \star} \mid \xi) \overset{iid}{\sim} \text{inv-Gamma} \left( \nu, \nu \xi \right), \quad \xi \sim \text{Gamma}(a_{\xi}, b_{\xi}),
\label{eq:priors}\end{equation}  
respectively, where values of ${\bm \mu}_{\beta}, {\bm \Sigma}_{\beta}, \nu, a_{\xi}$, and $b_{\xi}$ are previously set. Here inv-Gamma$(a, b)$ denotes the inverse gamma density with mean $b/(a-1)$, and Gamma$(a, b)$ is the gamma density with mean $b/a$. Other prior distributions are defined below.

\subsection{Partition} \label{ss:cl_modeling}

We establish a clustering model in the usual Bayesian nonparametric approach, wherein we specify a prior distribution for the random partition parameter. This prior distribution is formulated by using product partition models (PPM) introduced by \cite{Hartigan1990}. The main feature in this class of models is to express the prior distribution of the partition in a product form as $p\big( {\bm \rho}  = \{S_{1}, \ldots, S_{k}\} \big) \propto \prod_{j=1}^{k} C(S_{j})$, where $C(S_{j})$ is the cohesion function of $S_{j}$ and measures how likely the elements of $S_{j}$ are to co-cluster. Due to its flexibility in modeling heterogeneous data, this strategy has been used for different purposes. In the spatial context, \cite{Hegarty2008} employed a PPM to model partitions of areal units based on the number of its neighbours. This approach, called short boundary model, basically starts by defining the boundary length of the $i$th area, $\ell(i)$, as the total number of neighbors of area $i \in S_{j}$ not in component $S_{j}$. Then, the boundary length of a component $S_{j}$ is defined as $\ell(S_{j}) = \sum_{i\in S_j} \ell(i)$. This structure gives high probability to partitions into components having short boundary lengths. The cohesion function based on these concepts is formulated as:
\begin{equation}
    C(S_{j}) = \eta^{\ell(S_{j})},
\label{eq:cohesionHB} \end{equation} 
where $0 \leq \eta \leq 1$ determines how many components are in the partition in the following sense: when $\eta$ is small, most sampled partitions will consist of few large components, whereas for large values of $\eta$, they will consist of many small components. To avoid doubly intractable normalization constants, we keep this parameter fixed, as recommended by \cite{Hegarty2008}.

\subsection{Temporal components} \label{ss:time_modeling}

A very popular strategy in temporal modeling is to use autoregressive models, where the observed data at time $t$ is related to a number of lagged terms. Following this approach, we include a temporal structure in the model through a $q$-dimensional vector with lagged values of observed data. This vector could also include lagged terms related to seasonal information, depending on the context in which the model is applied. A key aspect of this specification is that it can be cast as a linear regression. The AR specification could just as easily be replaced by or complemented with harmonic functions. For this paper, we carried out exploratory studies in order to determine a good structure. We identified a patterned behavior of the series with a periodic component of 53 weeks, which can be seen in Figure~\ref{f:empirical_dta} of Appendix~\ref{appB}. We also compared different orders of autoregressive models and decided on order 3. Finally, our model includes one, two, three, and 53-weeks lagged values. The temporal dependence is defined for areal clusters induced by the PPM. Even though there is a single common cluster for all diseases, when considering a complete coefficient vector ${\bm \gamma}_{j}^{\star} = \{ {\bm \gamma}_{j1}^{\star}, \ldots, {\bm \gamma}_{jD}^{\star} \}$, the model admits that each disease behaves differently over time. This component is specified as:
\begin{equation}
    ({\bm \gamma}_{j}^{\star} \mid {\bm \mu}_{\gamma}, {\bm \Sigma}_{\gamma} ) \overset{iid}{\sim} \text{N}_{qD} \left({\bm \mu}_{\gamma}, {\bm \Sigma}_{\gamma} \right), \quad
    {\bm \mu}_{\gamma} \sim \text{N}_{qD} ({\bm \mu}_{\mu}, {\bm \Sigma}_{\mu}), \quad
    {\bm \Sigma}_{\gamma}\sim \text{inv-Wishart}(\text{df}, {\bm S}), 
\label{eq:gamma_prior} \end{equation}
where the hyperparameters ${\bm \mu}_{\mu}, {\bm \Sigma}_{\mu}, \text{df}$, and ${\bm S}$ are user-specified. 

\subsection{Spatial components} \label{ss:sp_modeling}

Although the cohesion function takes into account the number of neighbors of each area, it is restricted to a very specific setting and does not really impose spatial constraints. Trying to better explore spatial properties of our model, we include spatial random effects, $\bm \phi$, which are modeled using a class of directed acyclic graph autoregressive (DAGAR) models. Proposed by \cite{Datta2019}, DAGAR is formulated considering a predetermined ordering of the regions, ${\bm \pi}= \{\pi(1), \ldots, \pi(n)\}$, which is established by using a directed acyclic graph. Then, ${\bm \phi}_{d}$ is specified as a Gaussian distribution with zero mean and precision matrix ${\bm Q}(\alpha_{d}) = ({\bm I} - {\bm B})^{\top} {\bm \Lambda} ({\bm I} - {\bm B})$, where ${\bm B}$ is a $n \times n$ strictly lower-triangular matrix with elements $b_{ii'} = \frac{\alpha_{d}}{1+(n_{\pi(i)}-1)\alpha_{d}^{2}}$ if $i = 2, \ldots, n,$ and $i' \in N(i)$, otherwise, $b_{ii'} = 0$, where $n_{\pi(i)}$ is the cardinality of the neighbor set $N(i)$, and $0 \leq \alpha_{d} < 1$ represents the spatial correlation of disease $d$. The $n \times n$ diagonal matrix ${\bm \Lambda} = \text{diag}(\lambda_{i})$ is a $n \times n$ diagonal matrix with $\lambda_{i} = \frac{1+(n_{\pi(i)}-1)\alpha_{d}^{2}}{1-\alpha_{d}^{2}}$.  

Seeking to accommodate multivariate effects, \cite{Gao2022} extended DAGAR approach by defining a hierarchical structure that combines multiple univariate models. Basically, the spatial effect vector for the first disease is assumed to be DAGAR, i.e., ${\bm \phi}_{1} \sim \text{N}_{n} ({\bm 0}, \sigma^{2}_{\phi_{1}}{\bm Q}^{-1}(\alpha_{1}))$, where $\sigma^{2}_{\phi_{1}}$ is a scale parameter added to give more variability. Then, the conditional density of each ${\bm \phi}_{d}$ is progressively defined as ${\bm \phi}_{d} = {\bm A}_{d1}{\bm \phi}_{1} + \dots + {\bm A}_{d(d-1)}{\bm \phi}_{d-1} + {\bm \phi}_{d}$, for $d = 2, \ldots, D$, so that $({\bm \phi}_{d} \mid {\bm \phi}_{1}, \ldots, {\bm \phi}_{d-1}) \sim \text{N}_{n} ({\bm A}_{dd'} {\bm \phi}_{d'}, \sigma^{2}_{\phi_{d}}{\bm Q}^{-1}(\alpha_{d}))$. Each matrix ${\bm A}_{dd'} = \omega_{0dd'} {\bm I} + \omega_{1dd'} {\bm M}$ models the association between diseases $d$ and $d'$, with $\omega_{0dd'}$ being the coefficient that associates $\phi_{id}$ with $\phi_{id'}$, i.e., spatial effect of two diseases on the same area. Similarly, $\omega_{1dd'}$ associates $\phi_{id}$ and $\phi_{i'd'}$, i.e., spatial effect of two diseases on two different areas. ${\bm M}$ is the binary adjacency matrix for the map. Finally, the prior distribution of ${\bm \phi}$ can be written as: 
\begin{equation}
 ({\bm \phi} \mid \; {\bm \omega}, {\bm \alpha}, {\bm \sigma}^{2}_{\phi}) \sim \text{N}_{n} \Big({\bm \phi}_{1}; {\bm 0}, \sigma^{2}_{\phi_{1}} {\bm Q}^{-1}(\alpha_{1}) \Big) \prod_{d = 2}^{D} \text{N}_{n} \Big({\bm \phi}_{d}; {\bm A}_{dd'} {\bm \phi}_{d'}, \sigma^{2}_{\phi_{d}}{\bm Q}^{-1}(\alpha_{d}) \Big).
\label{eq:phiprior} \end{equation} 

MDAGAR formulation is completed by assuming prior distributions for ${\bm \omega}$, ${\bm \alpha}$, and ${\bm \sigma}^{2}_{\phi}$. In this case, we assume:
\begin{equation}
{\bm \omega}_{d} \sim \text{N}_{2}({\bm \mu}_{\omega}, {\bm \Sigma}_{\omega}), \quad \alpha_{d} \overset{iid}{\sim} \text{Beta}(a_{\alpha}, b_{\alpha}), \quad \sigma^{2}_{\phi_{d}} \overset{iid}{\sim} \text{inv-Gamma}(a_{\phi}, b_{\phi}).    
\label{eq:MDAGARpriors} \end{equation}

\subsection{Posterior inference}\label{ss:bayesiancomp}

We implement posterior simulation via MCMC, specifically, a hybrid Gibbs sampling that includes some Metropolis-Hastings moves. Let ${\bm \Omega} = \{ {\bm \beta}, {\bm \gamma}^{\star}, {\bm \mu}_{\gamma}, {\bm \Sigma}_{\gamma}, {\bm \rho}, {\bm \phi}, {\bm \omega}, {\bm \alpha}, {\bm \sigma}_{\phi}^{2}, {\bm \sigma}^{2 \star}, \xi \}$ denote the complete parameter vector. Thus, the joint posterior distribution is given by:
\begin{align}
    \nonumber p({\bm \Omega} \mid \; & {\bm Y}, {\bm X}) \propto \left[\prod_{j = 1}^{k} \prod_{i:i \in S_{j}} \prod_{t = q}^{T} \text{N} \left({\bm y}_{it}; X_{it}^{\top}{\bm \beta} + Z_{it}^{\top}{\bm \gamma}_{j}^{\star} + {\bm \phi}_{i}, {\bm \sigma}_{j}^{2 \star} \right) \right] \text{N}_{pD}\Big({\bm \beta}; {\bm \mu}_{\beta}, {\bm \Sigma}_{\beta} \Big) \\ 
   \nonumber & \times \left[ \prod_{j=1}^{k} \text{N}_{qD} \Big({\bm \gamma}_{j}^{\star}; {\bm \mu}_{\gamma}, {\bm \Sigma}_{\gamma} \Big) \right] \text{N}_{qD} \Big({\bm \mu}_{\gamma}; {\bm \mu}_{\mu},  {\bm \Sigma}_{\mu} \Big) \; \text{inv-Wishart} \Big({\bm \Sigma}_{\gamma}; \text{df}, {\bm S} \Big) \\ 
    \nonumber & \times \text{N}_{n} \Big({\bm \phi}_{1}; {\bm 0}, \sigma^{2}_{\phi_{1}} {\bm Q}^{-1}(\alpha_{1}) \Big) \left[ \prod_{d = 2}^{D} \text{N}_{n} \Big({\bm \phi}_{d}; {\bm A}_{dd'} {\bm \phi}_{d'}, \sigma^{2}_{\phi_{d}} {\bm Q}^{-1}(\alpha_{d}) \Big) \right] \\
    \nonumber & \times \left[ \prod_{d = 1}^{D} \text{N}_{D}({\bm \omega}_{d}; {\bm \mu}_{\omega}, {\bm \Sigma}_{\omega}) \; \text{Beta}(\alpha_{d}; a_{\alpha}, b_{\alpha}) \; \text{inv-Gamma} (\sigma^{2}_{\phi_{d}}; a_{\phi}, b_{\phi} ) \right] \\
    & \times \left[ \prod_{j=1}^{k} \prod_{d = 1}^{D} \text{inv-Gamma} \Big(\sigma_{jd}^{2 \star}; \nu, \nu \xi \Big) \right] \text{Gamma} \Big( \xi; a_{_{\xi}}, b_{_{\xi}} \Big).
\label{eq:posterior} \end{align} 

Although posterior inference is analytically intractable for this model, conditional conjugacy implies that some of the full conditionals are well-known distributions, which facilitates the sampling process. The most expensive update is the partition, where we use a strategy of auxiliary parameters based on \cite{Neal2000}’s algorithm number 8. The pseudo-code to sample from the posterior distribution \eqref{eq:posterior} is provided in Appendix~\ref{appA}.

\section{Simulation study} \label{s:simulations}

Throughout this section we detail two simulation studies that illustrate and highlight different aspects of the model previously presented. To do so, we choose the 70 microregions of the Brazilian state of Minas Gerais as our underlying map, where two regions are treated as neighbors if they share a common geographic boundary. Temporal trend is defined according to the goal of each study.

\subsection{Simulation 1: partition estimate} \label{ss:sim1}

The first simulation study is dedicated to explore the accuracy of partition estimates. To do so, we considered model \eqref{eq:model}--\eqref{eq:MDAGARpriors} with $D = 2$ as a data-generating mechanism to create 100 synthetic datasets. Two covariates were created and kept fixed for all simulations. One binary variable indicating high and low season across time and one continuous variable from N$(0, 0.5)$ independent across regions and times. Regression coefficients were set to ${\bm \beta}_{1} = (0.1, 0.4)^{\top}$ and ${\bm \beta}_{2} = (0.2, 0.3)^{\top}$, for diseases one and two, respectively. We adopted three different partition structures with $k = 2, 3$, and 4 clusters. Same variance value was set for all clusters and responses, $\sigma_{jd}^{2 \star} = 0.001$. Seeking to build data similar to the application context, we generated 120 time points whose trend was set using an AR(3) structure plus a 24-lagged value that represents a seasonal component. The cluster-specific autoregressive coefficients were set depending on the number of clusters as presented below:
\begin{equation*}
{\bm \gamma}_{d}^{\star \top} = \begin{bmatrix}
{\bm \gamma}_{1d}^{\star} \\
{\bm \gamma}_{2d}^{\star} \\
{\bm \gamma}_{3d}^{\star} \\
{\bm \gamma}_{4d}^{\star}
\end{bmatrix} = \begin{bmatrix}
0.5 & 0.3 & -0.5 & 0.4 \\
0.1 & -0.2 & -0.1 & 0.3 \\
1.6 & -0.9 & 0.1 & 0.1 \\
0.8 & 0.2 & -0.4 & 0.2 
\end{bmatrix}, \end{equation*}  
where each row represents one cluster, thus, $({\bm \gamma}_{1d}^{\star}, {\bm \gamma}_{2d}^{\star})$, $({\bm \gamma}_{1d}^{\star}, {\bm \gamma}_{2d}^{\star}, {\bm \gamma}_{3d}^{\star})$, and $({\bm \gamma}_{1d}^{\star}, {\bm \gamma}_{2d}^{\star}, {\bm \gamma}_{3d}^{\star}, {\bm \gamma}_{4d}^{\star})$ were considered for partitions with two, three, and four clusters, respectively. The values were the same for both responses. Regarding spatial random effects, we generated values of ${\bm \phi} = ({\bm \phi}_{1}^{\top}, {\bm \phi}_{2}^{\top})$ from a N$_{2}(\bm{0}, 10^{-5} {\bm V})$ distribution, where the precision matrix is:
\begin{equation*}
{\bm V}^{-1} = \begin{bmatrix}
{\bm Q}(\alpha_{1}) + {\bm A}_{21}^{\top} {\bm Q}(\alpha_{2}) {\bm A}_{21} & {\bm A}_{21}^{\top} {\bm Q}(\alpha_{2}) \\
{\bm Q}(\alpha_{2}) {\bm A}_{21} & {\bm Q}(\alpha_{2}) 
\end{bmatrix}, \end{equation*}
with $\alpha_{1} = \alpha_{2} = 0.5$, $Q(\alpha_{d}) = \alpha_{d}^{\#(i,i')}$, where $\#(i,i')$ represents the euclidean distance between a pair of areas, and ${\bm A}_{21}$ was defined with ${\bm \omega} = (1, 0.1)$.

We fit the model to the data simulated as indicated, considering some variations in prior distribution of the partition. For the cohesion function \eqref{eq:cohesionHB}, we let $\eta$ assume different values to evaluate how it impacts on partition estimates. In addition, we compare the results with the one obtained by using another popular cohesion function based on Dirichlet process, defined as $C(S_{j}) = M \; \Gamma(|S_{j}|)$ for some $M > 0$ \cite[see][for further details]{Quintana2003}. In this particular case, $M$ was set to 1. To each synthetic dataset we fit the model using the MCMC algorithm described in Appendix~\ref{appA} and save samples of size 1,000 that were obtained after running 20,000 iterates, discarding the first 50\% as burn-in and thinning by 10. Although the use of vague priors is generally indicated to carry out posterior inference, when it comes to models with complex structures it is practically convenient to use informative priors. Thus, we set the following hyperparameters: values in \eqref{eq:priors} are ${\bm \mu}_{\beta} = {\bm 0.25}$, ${\bm \Sigma}_{\beta} = \text{diag}({\bm 0.5})$, $\nu = 2$, $a_{\xi} = 1$, and $b_{\xi} = 2$, which represents a relatively noninformative prior given the data scale. Hyperparameters in the MDAGAR structure are defined to get more informative priors in order to avoid convergence issues. To do so, we set ${\bm \mu}_{\omega} = {\bm 0}$, ${\bm \Sigma}_{\omega} = \text{diag}({\bm 1})$ and $a_{\alpha} = b_{\alpha} = 300$ so that $\alpha_{d}$ is concentrated at 0.5 with small variance. For simplicity, we set $\sigma_{\phi}^{2} = 1$. The cluster-specific autoregressive coefficients also need to have an informative prior. Thus, after a preliminary analysis, we defined the values in \eqref{eq:gamma_prior} as ${\bm \mu}_{\mu} = {\bm 0}$, ${\bm \Sigma}_{\mu} = {\bm S} = \text{diag}({\bm 0.1})$, $\text{df} = 2(q + 1)$, where $q$ is the temporal design vector dimension, letting ${\bm \gamma}$ be relatively close to the truth. All partitions were estimated using the method available in the {\tt salso} R package \citep{Dahl2020} with the variation information (VI) loss function. To measure similarity between partitions, we used the Adjusted Rand Index (ARI, \cite{Hubert1985}) from the same {\tt salso} R package. The higher the ARI value, the closer the two clusterings are to each other. Furthermore, we used root mean squared error (RMSE) as a model fit performance indicator. 

\begin{figure} [H]
  \centerline{\includegraphics[width=\textwidth]{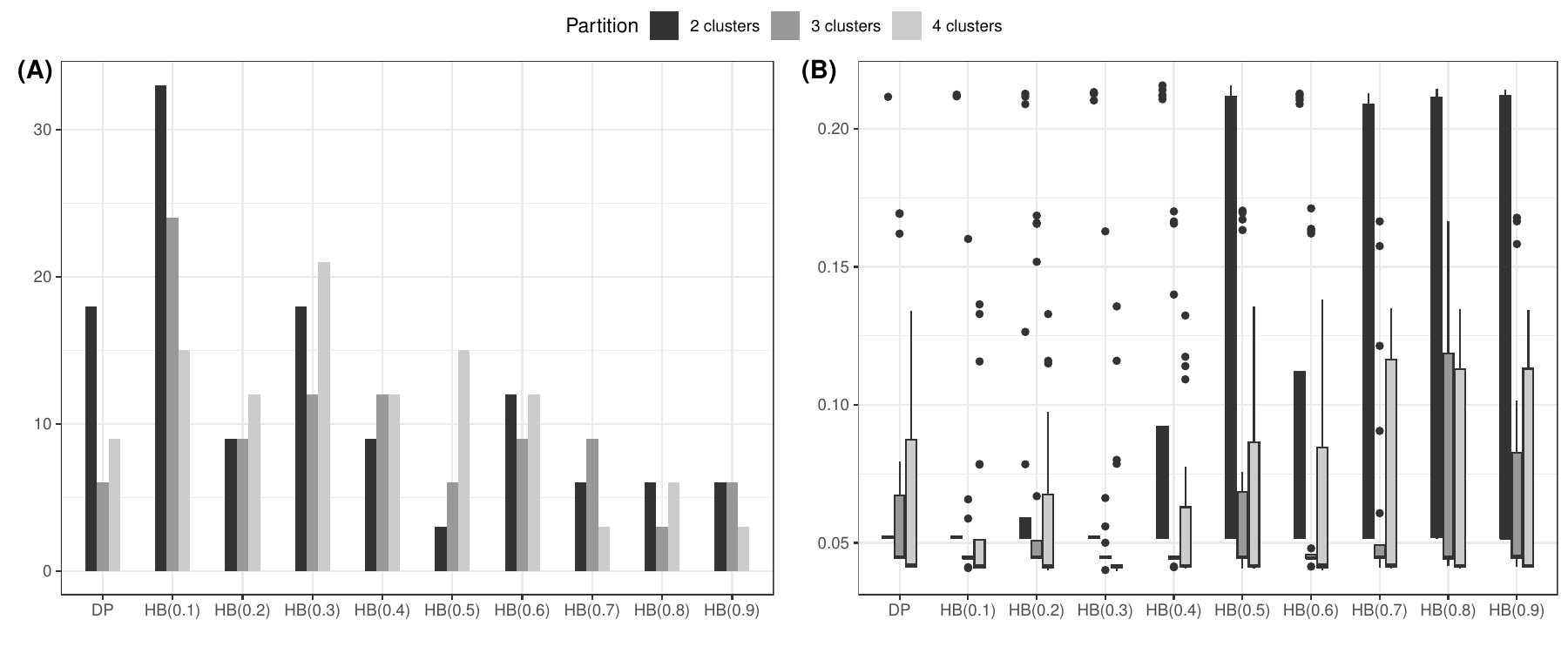}}
  \caption{Performance summary measures of the 10 models for the three scenarios. (A): Barplot indicating how frequently each model obtained the highest ARI value over 100 datasets. In case of a tie, both models were credited. (B): Boxplot of RMSE values.
  }
\label{fig:study1}
\end{figure}

Performance summary measures of the 10 model configurations (PPM-DP and PPM-HB with $\eta \in \{0.1, 0.2, \ldots, 0.9 \}$) for the three scenarios (2, 3, and 4 clusters) are displayed in Figure~\ref{fig:study1}. In terms of partition estimation, Figure~\ref{fig:study1}(A) indicates how frequently each model obtained the highest ARI value over the 100 datasets. PPM-HB($\eta = 0.1$) model showed better estimation capacity in scenarios with $k = 2, 3$, while PPM-HB($\eta = 0.3$) was the one that performed best for $k = 4$. Specifically, when there are only two clusters, the model estimated the partition correctly in most datasets. However, as the number of clusters increased, the accuracy of partition estimates decreased. ARI values, averaged over the 100 datasets, were 0.89, 0.66, and 0.54 for $k = 2, 3, 4$, respectively. In addition, lower RMSE values were obtained by models with lower values to the parameter in the HB cohesion function; see Figure~\ref{fig:study1}(B). Indeed, \cite{Hegarty2008} had already warned that the clustering is more accurate for $0.1 < \eta < 0.5$. Outside this range, the model may either not perform well in terms of clustering, and even lead to estimation with many clusters. 

\subsection{Simulation 2: modeling performance} \label{ss:sim2}

The second simulation study is designed to compare our model to the one implemented in {\tt CARBayesST} R package \citep{Lee2023}. In particular, we used the {\tt MVST.CARar} function, where the single set of random effects is modeled as jointly correlated over time, space, and outcome. We generated 100 synthetic datasets from model \eqref{eq:model}--\eqref{eq:MDAGARpriors} with $D = 2$. For fairer comparison, we respected the restrictions of the competing model and considered an AR(2) and a partition structure with only one cluster. Thus, we generated 120 time points, of which we used 100 for model estimation and kept 20 to validate the prediction. In addition, our study also contemplates other scenarios with two and three clusters to check the performance of our model. Covariates, regression coefficients, spatial effects, and variance were generated the same as the first simulation study (Section~\ref{ss:sim1}). Cluster-specific autoregressive coefficients were set depending on the number of clusters as shown below:
\begin{equation*}
{\bm \gamma}^{\star \top} = \begin{bmatrix}
{\bm \gamma}_{1d}^{\star} \\
{\bm \gamma}_{2d}^{\star} \\
{\bm \gamma}_{3d}^{\star} 
\end{bmatrix} = \begin{bmatrix}
1.6 & -0.7  \\
0.9 & -0.1 \\
0.3 & 0.1
\end{bmatrix}. \end{equation*}

As before, each row represents one cluster, so, ${\bm \gamma}_{1d}^{\star}$, $({\bm \gamma}_{1d}^{\star}, {\bm \gamma}_{2d}^{\star})$, and $({\bm \gamma}_{1d}^{\star}, {\bm \gamma}_{2d}^{\star}, {\bm \gamma}_{3d}^{\star})$ were considered for $k = 1, 2, 3$, respectively. Both responses were generated with the same temporal coefficients. To each synthetic dataset we fit our proposed model considering PPM-DP and PPM-HB($\eta = 0.35$), and the competing model. We save samples of size 1,000 that were obtained after running 15,000 iterates, discarding the first 5,000 as burn-in and thinning by 10. Most hyperparameters were defined as in Section~\ref{ss:sim1}, with the exception of ${\bm \mu}_{\mu}$, ${\bm \Sigma}_{\mu}$, and ${\bm S}$. As previously mentioned, it is preferable to specify a rather informative prior distribution for this cluster structure. However, considering that the model in this study has a simpler temporal structure than the first, we can provide a little more variability to these hyperparameters. We thus set now ${\bm \mu}_{\mu} = (1,0,1,0)$ and ${\bm \Sigma}_{\mu} = {\bm S} = \text{diag}({\bm 0.5})$. Results from this simulation study are presented in Table~\ref{tab:study2}. In addition to log-likelihood and RMSE, we used different criteria as model fit performance indicators \citep{Gelman2014}. Akaike information criterion (AIC) and Bayesian information criterion (BIC), which consider the number of fitted parameters as a parsimony penalty term, indicated that PPM-DP and PPM-HB perform better than our competitor. On the other hand, deviance information criterion (DIC), and Watanabe-Akaike information criterion (WAIC), which use the effective number of parameters as bias correction, pointed out {\tt MVST.CARar} as the best model. PPM-HB was slightly superior to PPM-DP in all scenarios according to all criteria. 

\begin{table}[H]\centering
 \def\~{\hphantom{0}}
 \begin{minipage}{\textwidth}
 \caption{Model fit performance metrics. The bold font identifies best model fits to each scenario in terms of each criterion. Higher values for log-likelihood indicate better fit while lower values for RMSE, AIC, BIC, DIC, and WAIC indicate better fit.}
\label{tab:study2}
  \begin{tabular*}{\textwidth}{@{}l@{\extracolsep{\fill}}c@{\extracolsep{\fill}}c@{\extracolsep{\fill}}c@{\extracolsep{\fill}}c@{\extracolsep{\fill}}c@{\extracolsep{\fill}}c@{\extracolsep{\fill}}c@{\extracolsep{\fill}}c@{\extracolsep{\fill}}c@{\extracolsep{\fill}}c@{\extracolsep{\fill}}c@{\extracolsep{\fill}}c@{}}
  \hline
& & \multicolumn{3}{c}{{1 cluster}} & & \multicolumn{3}{c}{{2 clusters}} & & \multicolumn{3}{c}{{3 clusters}} \\ [1pt]
\cline{3-5} \cline{7-9}  \cline{11-13} \\ [-6pt]
 & & DP & HB & CAR & & DP & HB & CAR & & DP & HB & CAR \\ 
 \hline
 Log-likelihood            & & 26029 & 27981 & {\bf 35660}    & & 16391 & 21006 & {\bf 30877}    & & 10361 & 16189 & {\bf 34098} \\
 RMSE                      & & 0.07 & 0.03 & {\bf 0.02}       & & 0.18 & 0.05 & {\bf 0.04}       & & 0.31 & 0.08 & {\bf 0.02} \\
 p$_{_\text{\tiny total}}$ & & 154 & 154 & 14012              & & 157 & 154 & 14012              & & 162 & 159 & 14012 \\
 AIC                       & & -51750 & {\bf -55654} & -43296 & & -32468 & {\bf -41704} & -33729 & & -20397 & {\bf -40173} & -32060 \\
 BIC                       & & -50585 & {\bf -54492} & 62450  & & -31287 & {\bf -40542} & 72017  & & -19176 & {\bf -30862} & 65573 \\
 p$_{_\text{\tiny DIC}}$   & & -781 & 150 & 5872              & & 28 & 95 & 4306                 & & 125 & 159 & 6831 \\
 DIC                       & & -52840 & -55813 & {\bf -59577} & & -32753 & -41917 & {\bf -53141} & & -20596 & -32326 & {\bf -54535} \\
 p$_{_\text{\tiny WAIC}}$  & & 1345 & 149 & 4867              & & 403 & 163 & 5716               & & 359 & 212 & 3987 \\
 WAIC                      & & -50460 & -55811 & {\bf -59118} & & -32315 & -41843 & {\bf -50220} & & -20377 & -32154 & {\bf -56224} \\
\hline
\end{tabular*} \end{minipage} \vspace{10pt}
\end{table} 

\begin{figure} [H]
  \centerline{\includegraphics[width=\textwidth]{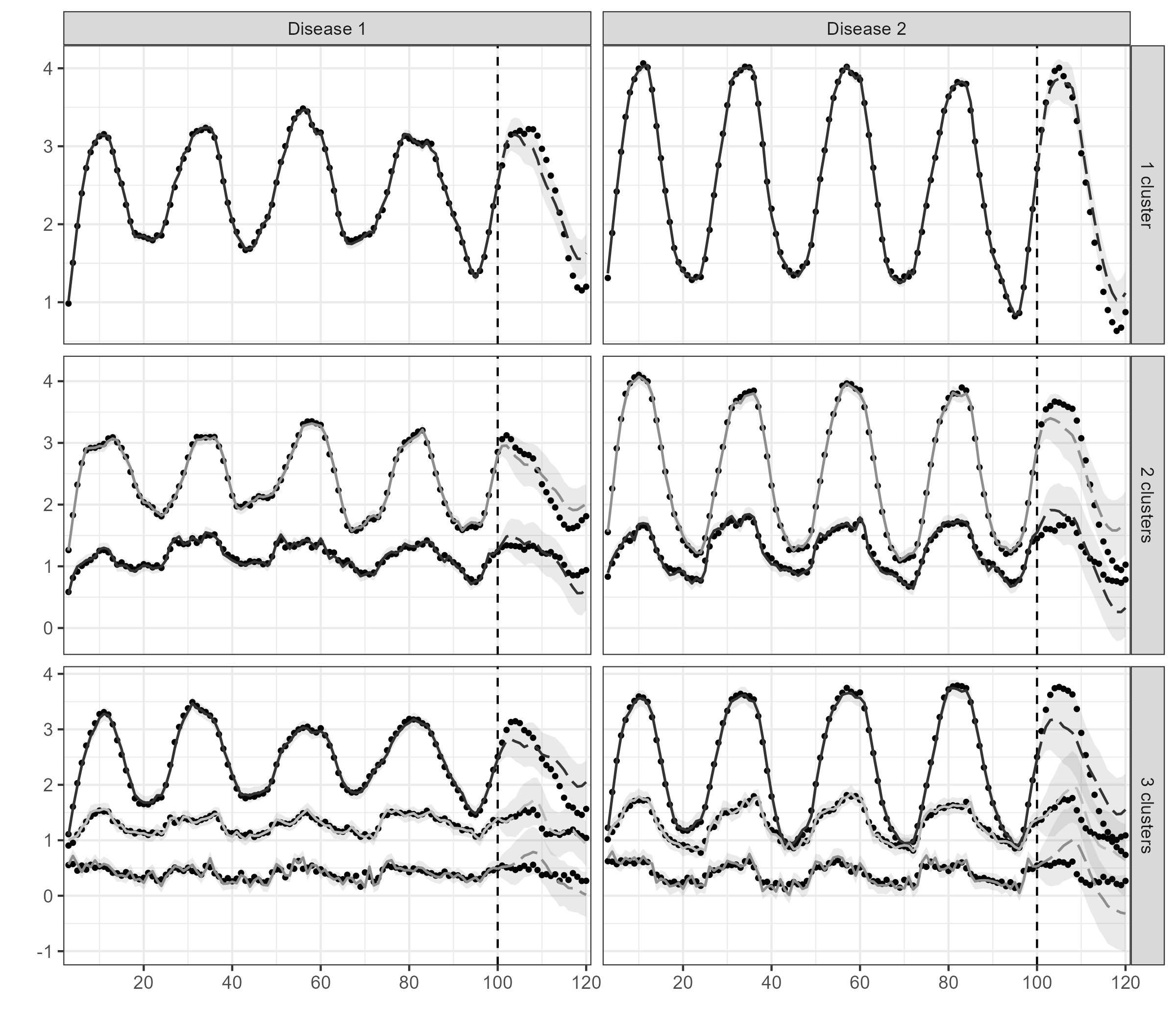}}
  \caption{Observed data (points), estimated time series (solid lines), and 20-points prediction (dashed lines) of both response for one area of each estimated cluster within of three scenarios (with one, two, and three clusters). Estimation and prediction values are accompanied by their respective 95\% credible intervals.}
\label{fig:study2_series}
\end{figure}

To study the ability to recover regression coefficients, we checked mean and 95\% credible intervals for $\bm \beta$. We identified that model PPM-DP and PPM-HB estimated coefficients much closer to the truth than the competitor, which is repeated for all scenarios. Furthermore, PPM-HB was superior to PPM-DP. In Figure~\ref{f:beta_coverage} of Appendix~\ref{appB}, we present a summary of $\beta$'s estimated by the three models under each scenario. Ultimately, we checked the predictive ability of PPM-HB. Figure~\ref{fig:study2_series} displays the estimated time series, a 20-point prediction, and their respective 95\% credible intervals for one chosen area for each estimated cluster under every scenario. Even though our model has a reduced number of parameters and random effects compared to the competitor, it proved to be efficient in terms of estimation and prediction. 

The upshot from this simulation study is that the PPM model either with DP or HB cohesion function performs well in terms of recovering regression coefficients, estimation of time series, and prediction of new data points. As seen in the simulation study presented in Section~\ref{ss:sim1}, PPM-HB estimates the partition better than PPM-DP, which reverberates in the model fitting. The performance of our competing model is satisfactory in terms of fitting. Nevertheless, the large number of parameters and random effects could lead to overfitting, in addition to the elevated computational cost.

\section{Application to mosquito-borne disease data} \label{s:dataapp}

In this section we describe application of the proposed model to mosquito-borne diseases, specifically dengue and chikungunya cases in the Brazilian Southeast region, as described in Section~\ref{s:motivation}. We performed posterior inference for model \eqref{eq:model}--\eqref{eq:MDAGARpriors} from a sample of size 1,000, by running 20,000 iterations, discarding the first 50\% as burn-in, and thinning by 10. Convergence was monitored graphically. To provide a point estimate of the random partition, we minimized the VI loss function with identical misclassification cost parameters. As in the simulation study, calculations were done via {\tt salso} R package. RMSE was used as model fit performance indicator for prior elicitation and for choosing the order of diseases in the MDAGAR structure. In addition to the analysis where the two diseases are jointly modeled, we perform independent univariate analyzes so that we can compare results. All information about the univariate analysis is presented in Appendix~\ref{appC}. 

Considering the model complexity and high-dimensionality of the dataset, prior elicitation was carefully carried out through a preliminary analysis (not shown). Hyperparameters related to the variance ($\sigma_{jd}^{2 \star}$) and MDAGAR structure were kept the same as those set in the simulation studies (see Section~\ref{s:simulations}), with the inclusion of $a_{\phi} = 2$, $b_{\phi} = 0.1$ so that $\sigma^{2}_{\phi_{d}}$ assumes small values. ${\bm \mu}_{\beta} = {\bm 0}$ and ${\bm \Sigma}_{\beta} = \text{diag}({\bm 1})$ represent a relatively noninformative prior to regression coefficients. Recall that, for computational convenience, the hyperparameters to cluster-specific autoregressive coefficients should be cautiously set. In this case, the convergence was achieved with ${\bm \mu}_{\mu} = {\bm 0}$, ${\bm \Sigma}_{\mu} = \text{diag}({\bm 1})$, $\text{df} = 2(q + 1)$, and ${\bm S} =  nT \tilde{\sigma}^{2} (Z^{\top}Z)^{-1}$, where $q$ is the temporal design vector dimension, $n$ and $T$ are the number of areas and times, respectively, and $\tilde{\sigma}^{2}$ is the estimated variance of $\gamma$ via ordinary least squares. Regarding $\eta$, we explored a range of values and ended up setting $\eta = 0.35$.

\begin{figure}[H]
 \centerline{\includegraphics[width=4in]{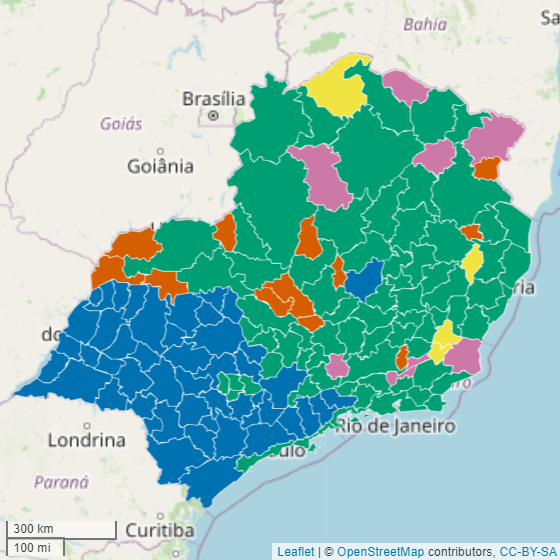}}
\caption{Posterior estimate of the random partition to the Brazilian Southeast region obtained by minimizing the variation of information loss function with identical cost parameters for misclassification.}
\label{f:est_part}
\end{figure}

Our model estimated a partition with five clusters differentiated by color in Figure~\ref{f:est_part}. The first cluster ($S_{1}$, blue, 49 areas) is predominantly made up of areas of S{\~a}o Paulo state, but it also includes the state capital of Minas Gerais, Belo Horizonte. Areas belonging to $S_{1}$ are basically characterized by the prevalence of dengue fever (reaching almost 30 thousand cases in Belo Horizonte in the last week of April 2019) and low occurrence of chikungunya cases (95 cases in a week maximum). Another aspect of this cluster is its high population density. Areas from the second cluster ($S_{2}$, green, 72 areas) also presented prevalence of dengue fever, however, different from $S_{1}$, outbreaks of chikungunya were identified throughout the period. The biggest spikes in dengue and chikungunya were registered in the city of Sete Lagoas - Minas Gerais in the same week, mid-March 2021. The third and smallest cluster ($S_{3}$, yellow, 4 areas) is the only one where there was prevalence of chikungunya. The fourth cluster ($S_{4}$, purple, 8 areas) is probably the cluster with the greatest balance between diseases. On average, 15 cases of dengue and 13 of chikungunya were reported weekly. The last cluster ($S_{5}$, orange, 12 areas) is similar to $S_{1}$. Very few cases of chikungunya were reported in these areas, totaling 937 for the entire observed period. On the other hand, the highest dengue rates were observed in areas belonging to this cluster. It is worth mentioning that unlike $S_{1}$, $S_{5}$ is formed by low-populated areas, 92 thousand inhabitants on average. In Figure~\ref{f:empirical_dta} of Appendix~\ref{appB}, we present empirical information on data that helps to understand the cluster formation.

When comparing estimated partitions obtained from bivariate and univariate analysis, Figure~\ref{f:est_part} and Figure~\ref{f:est_part_app} of Appendix~\ref{appC}, respectively, we can identify some similarities. For example, $S_{1}$ is also seen in the univariate estimated partitions with only few variations. Even so, in this case we can say that the cluster formed for the chikungunya data prevails in the bivariate case since they differ only by one area, Pouso Alegre (SP). In general, the partition obtained for chikungunya is more fragmented than other two, with 8 clusters versus 5 found in other cases. This is because while dengue is a more widely spread disease in Brazil, chikungunya is concentrated in some areas and periods.

Posterior mean of cluster-specific parameters conditioned on estimated partition are displayed in Figure~\ref{f:cl_parameters} of Appendix~\ref{appB}. Coefficients for one and two-weeks lagged observed data values, respectively, have a positive effect on the number of cases for all clusters and diseases. Contrarily, the coefficient related to seasonal information has a negative effect. Finally, the coefficient for three-weeks lagged values has a negative effect on all dengue clusters, while in case of chikungunya, it has a negative effect on $S_{3}$ and close to zero for the other clusters. In general, cases of both diseases depend on cases from the previous two weeks. However, the dependence on cases of more lagged weeks differs between diseases, which may indicate different temporal patterns. The estimated covariance matrix ${\bm \Sigma}_{\gamma}$ corroborates the existence of dependence between infections (see Appendix~\ref{appB}). Variance parameters are generally concentrated on low values, but they also proved to be different across clusters and diseases. 

Regarding regression coefficients (Figure~\ref{f:cl_parameters} of Appendix~\ref{appB}), as expected, summer and autumn have a positive effect on the number of cases. This occurs because in this Brazilian region, summer is a rainy and warm season, which are favorable characteristics for the development of mosquitoes. The beginning of autumn is still marked by rain and high temperatures providing appropriate conditions for mosquito incubation, and then, the spread of diseases. This is exactly the opposite in winter, characterized by drought and low temperatures, which also impacts in spring. This behavior was observed for both diseases, although values for dengue were a bit higher. In turn, HDI exhibited the opposite effect for each disease. The posterior average of this parameter is positive for dengue, but negative for chikungunya.

From MDAGAR construction, it is clear that the order of diseases may lead to different results, so it is convenient to compare all permutations. In this case, we used RMSE to compare the two possible combinations, dengue given chikungunya and chikungunya given dengue, and the first was chosen. Thus, all presented results were obtained from the model with dengue given chikungunya. Recall that, in this specification, $\alpha_{1}$ represents the residual spatial autocorrelation for dengue after accounting for the explanatory variables, $\alpha_{2}$ is the residual spatial autocorrelations after accounting for explanatory variables and dengue, and ${\bm \omega}$ reflects the associations between diseases. In this case, both infections exhibited moderate spatial autocorrelation (0.66 [0.61, 0.72] for dengue and 0.50 [0.46, 0.54] for chikungunya). Furthermore, the variability of dengue spatial random effects is larger than chikungunya (2.63 [1.98, 3.43] and 0.02 [0.01, 0.02], respectively). Finally, we found that these arboviruses are positively associated within an area (0.99 [0.97, 1.01]), which does not happen between neighbors (0.00 [0.00, 0.01]). 

\begin{figure}
  \centerline{\includegraphics[width=0.97\textwidth]{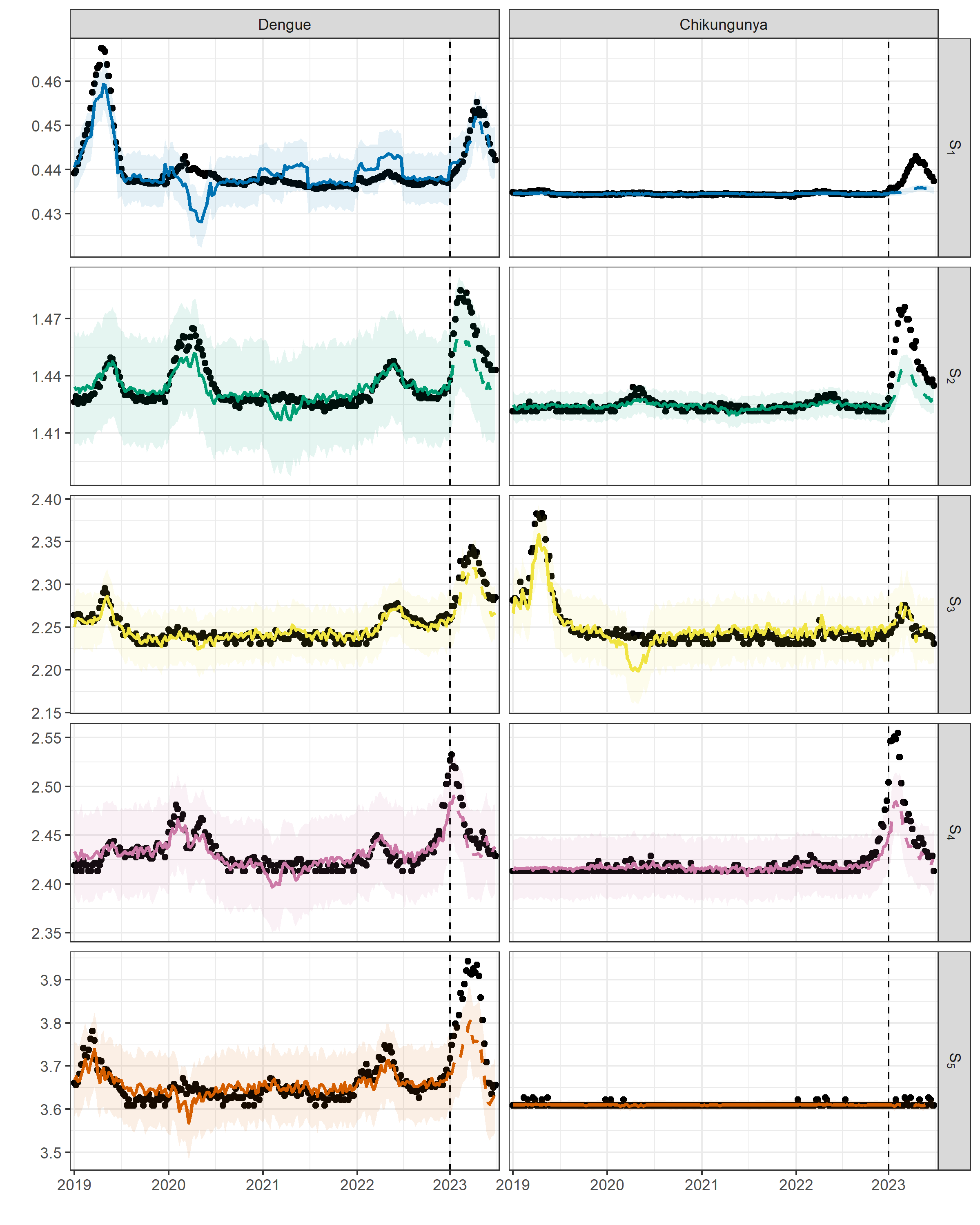}}
  \caption{Observed data (points), estimated time series (solid lines), and 26-points prediction (dashed lines) of dengue fever and chikungunya for one area of each estimated cluster. Estimation and predication are accompanied by their respective 95\% credible intervals. From up to down: Belo Horizonde - MG, Teofilo Otoni - MG, Itaperuma - RJ, Almenara - MG, and Monte Carmelo - MG.}
\label{fig:ts_clusters}
\end{figure}

Lastly, Figure~\ref{fig:ts_clusters} displays the estimated time series and a 26-point prediction (i.e., 6 months) of both responses for one area of each estimated cluster. Overall, our model proved to be efficient in terms of estimation. Nevertheless, it was not able to predict peaks as high as those that occurred in 2023, especially in cities where there were no serious outbreaks observed during the years with available data. According to the World Health Organization, in 2023, the Americas have seen a sharp increase in mosquito-borne disease cases, characterized by a significant increase in the number, scale, and simultaneous occurrence of multiple outbreaks, spreading into regions previously unaffected. Incorporating data from 2023 into the model may improve its predictive capacity, by explicitly considering this last relevant outbreak.

\section{Conclusion} \label{s:conclusion}

Motivated by the peculiarities of mosquito-borne diseases data and as a contribution to the literature on flexible approaches for this context, we developed a spatio-temporal model where the temporal dependence is defined for areal clusters induced by product partition models. The versatility of the temporal component in this approach lies in its ability to incorporate any order autoregressive models or even harmonic functions. Furthermore, it is possible to include components that account for potential seasonal patterns, which can be highly beneficial when modeling arboviruses. Coefficients related to the temporal structure are disease and cluster-specific. Additionally, our model allows for the inclusion of spatial, temporal, and/or spatio-temporal covariates, whose coefficients are also specific for each response. The neighboring dependence is also incorporated into the model by way of a directed acyclic graph autoregressive structure, that provides us with spatial correlation parameters for each disease, in addition to parameters that relate them. 

Through simulation studies, we showed that our methodology is capable of establishing spatial partitions for areas where the temporal trends are similar. Furthermore, we compared our results with a competing model and found that our approach provides good estimation performance in addition to a good predictive capacity. Finally, we applied the proposed model to a Brazilian dataset of dengue and chikungunya cases per areal unit and epidemiological week. The analysis allowed us: (i) to identify spatial clusters that are characterized by differences in the temporal trend, (ii) to observe different spatial and temporal behavior for each disease, (iii) to understand how diseases correlate with each other, and (iv) to obtain a 6 month prediction. Additionally, we provided results obtained from independent disease analysis, which allows us to make comparisons. 

Performing posterior simulation via MCMC in random partition models is computationally intensive, and our study did not escape this ambush unscathed. Hence, future work includes the exploration of alternative strategies to overcome this limitation, for instance, reducing the number of partitions that compose the search space, or the implementation of different posterior approximation strategies. Furthermore, our model can be extended to other distribution families in order to accommodate data that exhibit different features.

\section*{Acknowledgements}

This paper was partially supported by grant FONDECYT 1220017.

\bibliographystyle{biom} \bibliography{reference}

\appendix
\section*{Appendix}
\section{MCMC algorithm} \label{appA}

In this section, we describe the Metropolis-within-Gibbs sampler algorithm used to obtain posterior samples from the joint distribution presented in Section~\ref{ss:bayesiancomp}. Same as before, we use ${\bm \Omega} = \{ {\bm \beta}, {\bm \gamma}^{\star}, {\bm \mu}_{\gamma}, {\bm \Sigma}_{\gamma}, {\bm \rho}, {\bm \phi}, {\bm \omega}, {\bm \alpha}, {\bm \sigma}_{\phi}^{2}, {\bm \sigma}^{2 \star}, \xi \}$ to denote the complete parameter vector, then the algorithm is given as follow:

\begin{enumerate}
    \item Update $\xi$:
\begin{equation*}
    \xi \mid \cdot \sim \text{Gamma}\left( a_{\xi}, \; \sum\limits_{j=1}^{k} \sum_{d=1}^{D} \frac{\nu}{\sigma_{jd}^{2 \star}} + b_{\xi} \right).
\end{equation*}

    \vspace{0.5cm} \item Update ${\bm \mu}_{\gamma}$:
\begin{equation*}
    {\bm \mu}_{\gamma} \mid \cdot \sim \text{N} \left( \left( k \Sigma_{\gamma}^{-1} + \Sigma_{\mu}^{-1}\right)^{-1}\left( \sum_{j=1}^{k} {\bm \gamma}_{j}^{\star} \Sigma_{\gamma}^{-1} + {\bm \mu}_{\mu} \Sigma_{\mu}^{-1} \right), \; \left( k \Sigma_{\gamma}^{-1} + \Sigma_{\mu}^{-1}\right)^{-1} \right).
\end{equation*}

   \vspace{0.5cm} \item Update ${\bm \Sigma}_{\gamma}$:
\begin{equation*}
    \Sigma_{\gamma} \mid \cdot \sim \text{inv-Wishart} \left( \text{df} + k, \; S + \sum_{j=1}^{k} ({\bm \gamma}_{j}^{\star} - {\bm \mu}_{\gamma})^{\top}({\bm \gamma}_{j}^{\star} - {\bm \mu}_{\gamma}) \right).
\end{equation*}

    \vspace{0.5cm} \item Update $\rho$: the following update is based on Algorithm 8 of \cite{Neal2000}. To do so, let $k^{-}$ denote the number of clusters after removing the $i-$th area from the sample and $S_{j}^{-}$ the corresponding cluster. Then, for each $i = 1, \ldots, n$, sample from:
\begin{equation*}
    P[c_{i} = j \mid {\bm c}_{-i}, {\bm Y},  {\bm \Omega} ] \propto \begin{cases} 
\frac{C(S_{j}^{-} \cup \; \{i \})}{C(S_{j}^{-})} \prod\limits_{t = 1}^{T} \mathcal{L}({\bm \Omega}_{i}), \quad \text{if} \; j = 1, \ldots, k^{-} \\
C(\{i \}) \prod\limits_{t = 1}^{T} \mathcal{L}({\bm \Omega}_{i}), \quad \text{if} \; j = k^{-} + 1, \end{cases}
\end{equation*}
where $\mathcal{L}$ denotes the likelihood function of ${\bm \Omega}_{i}$, in this case $\text{N}(Y_{it}; X_{it}^{\top}{\bm \beta} + Z_{it}^{\top}{\bm \gamma}_{j}^{\star} + \phi_{i}, \, \sigma_{j}^{2 \star})$. Cohesion function is defined as $C(S_{j}) = \eta^{\ell(S_{j})}$ for \cite{Hegarty2008}'s specification and $C(S_{j}) = M\times \Gamma(|S_{j}|)$ when based on the Dirichlet process.

    \vspace{0.5cm} \item Update $\sigma_{jd}^{2 \star}$: for each $j = 1, \ldots, k$ and $d = 1, \ldots, D$, sample from:
\begin{equation*}
    \sigma_{jd}^{2 \star} \mid \cdot \sim \text{inv-Gamma}\left( \frac{n_{j}(T-q)}{2} + \nu, \; \sum_{i=1}^{n} \sum_{t=q}^{T} \frac{ \big( y_{itd} - {\bm X}_{itd}^{\top} {\bm \beta}_{d} - {\bm Z}_{itd}^{\top} {\bm \gamma}_{jd}^{\star} - {\bm \phi}_{id} \big)^{2}}{2} + \nu \xi \right),
\end{equation*}
where $n_{j}$ is the number of areas belonging to cluster $j$.

    \vspace{0.5cm} \item Update ${\bm \gamma}_{j}^{\star}$: for each $j = 1, \ldots, k$, sample from:
\begin{equation*}
    {\bm \gamma}_{j}^{\star} \mid \cdot \sim \text{N} \left(m_{\gamma}, V_{\gamma}\right),
\end{equation*}
where $V_{\gamma} = \left( \sum\limits_{i:i \in S_{j}} \sum\limits_{t=q}^{T} \frac{ Z_{it}Z_{it}^{\top}}{\sigma_{j}^{2 \star}} + \Sigma^{-1}_{\gamma} \right)^{-1}$ and $m_{\gamma} = V_{\gamma} \left( \sum\limits_{i:i \in S_{j}} \sum\limits_{t=q}^{T} \frac{ (y_{it} - X_{it}^{\top}{\bm \beta} - {\bm \phi}_{i}) Z_{it}^{\top}}{\sigma_{j}^{2 \star}} + \Sigma^{-1}_{\gamma} \mu_{\gamma} \right)$.
    
    \vspace{0.5cm} \item Update ${\bm \omega}_{d}$: to extract ${\bm \omega}_{d}$ from the matrix ${\bm A}$ used in multivariate DAGAR construction, we can rewritten ${\bm A}_{dd'}{\bm \phi}_{d'} = {\bm W}_{d'} {\bm \omega}_{dd'}$ where ${\bm W}_{d'} = ({\bm \phi}_{d'}, {\bm \epsilon}_{d'})$ and ${\bm \epsilon}_{d'} = \left( \sum\limits_{i' \sim 1} \phi_{d'i'}, \ldots, \sum\limits_{i' \sim n} \phi_{d'i'} \right)^{\top}$. Consequently, ${\bm \phi}_{d} = {\bm \delta}_{d} {\bm \omega}_{d} + {\bm \epsilon}_{d}$, where ${\bm \delta}_{d}$ is a block matrix so that ${\bm \delta}_{d} = (W_{1}, \ldots, W_{d - 1})$. Finally, for each $d = 2, \ldots, D$, sample from:
\vspace{0.2cm} \begin{equation*}
    {\bm \omega}_{d} \mid \cdot \sim \text{N} ( m_{\omega}, V_{\omega} ) \\
\end{equation*}
where $V_{\omega} = \left({\bm \delta}_{d}^{\top} \frac{{\bm Q}(\alpha_{d})}{\sigma_{\phi_{d}}^{2}} {\bm \delta}_{d} + \Sigma_{\omega}^{-1} \right) ^{-1}$ and $m_{\omega} = V_{\omega} \left({\bm \delta}_{d}^{\top} \frac{{\bm Q}(\alpha_{d})}{\sigma_{\phi_{d}}^{2}} {\bm \phi}_{d} + \Sigma_{\omega}^{-1} \mu_{\omega} \right)$.
    
    \vspace{0.5cm} \item Update $\sigma_{\phi_{d}}^{2}$: for each $d = 1, \ldots, D$, sample from:
\begin{equation*}
    \sigma_{\phi_{d}}^{2} \mid \cdot \sim \text{inv-Gamma} \left( a_{\phi} + \frac{n}{2}, \bar{b}_{\phi} \right),
\end{equation*}
where 
\begin{equation*}
\bar{b}_{\phi} = \begin{cases} 
b_{\alpha} + \frac{1}{2} {\bm \phi}_{1}^{\top} {\bm Q}(\alpha_{1}) {\bm \phi}_{1}, \quad \text{if} \; d = 1 \\
b_{\alpha} + \frac{1}{2} \left({\bm \phi}_{d} - \sum\limits_{d' = 1}^{d-1} {\bm A}_{dd'} {\bm \phi}_{d'} \right)^\top {\bm Q}(\alpha_{d}) \left({\bm \phi}_{d} - \sum\limits_{d' = 1}^{d-1} {\bm A}_{dd'} {\bm \phi}_{d'}\right), \quad \text{if} \; d = 2, \ldots, D. \end{cases}
\end{equation*}

    \vspace{0.5cm} \item Update $\alpha_{d}$: the following update includes an adaptive random walk Metropolis step as presented by \citet[Chapter 8]{Givens2012}. Besides, it is performed over the transformed variable, hence, let ${\bar \alpha}_{d} = \log \left( \frac{\alpha_{d}}{1 - \alpha_{d}} \right)$, then the full conditional distribution of ${\bar \alpha}_{d}$ is given by: 
\vspace{0.2cm} \begin{equation*}
    p({\bar \alpha}_{d} \mid ) \propto p({\bm \phi}_{d} \mid {\bm \omega}_{d}, \sigma_{\phi_{d}}^{2}) \times p(\alpha_{d}) \mid J \mid,
\end{equation*} \vspace{0.2cm}

where the Jacobian is $J = \frac{\alpha_{d}}{1 - \alpha_{d}}$ while prior distributions $p({\bm \phi}_{d} \mid {\bm \omega}_{d}, \sigma_{\phi_{d}}^{2})$ and $p(\alpha_{d})$ are defined by (5) and (6), respectively.

    \vspace{0.5cm} \item Update ${\bm \phi}_{d}$: for each $d = 1, \ldots, D$, sample from:
\begin{equation*}
    {\bm \phi}_{d} \mid \cdot \sim \text{N} \Big( m_{d}, V_{d} \Big)
\end{equation*}
where
\begin{equation*}
V_{d}  = \begin{cases} 
\left( \frac{{\bm Q}(\alpha_{d})}{\sigma_{\phi_{d}}^{2}} + \sum\limits_{l = d + 1}^{D} A_{ld}^{\top} \frac{{\bm Q}(\alpha_{l})}{\sigma_{\phi_{l}}^{2}} A_{ld} + \frac{1}{\sigma_{d}^{2}} I_{n} \right)^{-1}, \quad \text{if} \; d = 1, \ldots, D - 1 \\
\left(\frac{{\bm Q}(\alpha_{d})}{\sigma_{\phi_{d}}^{2}} + \frac{1}{\sigma_{d}^{2}} I_{n} \right)^{-1}, \quad \text{if} \; d = D \end{cases} 
\end{equation*}
and
\begin{align*}
m_{d} = 
\begin{cases} 
V_{d} \left[ A_{21}^{\top} \frac{Q(\alpha_{2})}{\sigma_{\phi_{2}}^{2}} \phi_{2} + \sum\limits_{l = 3}^{D} A_{l1}^{\top} \frac{Q(\alpha_{l})}{\sigma_{\phi_{l}}^{2}} \left( {\bm \phi}_{l} - \sum\limits_{l' = 2}^{l-1} A_{ll'} {\bm \phi}_{l'} \right) + \frac{1}{\sigma_{1}^{2}} ({\bm Y}_{1} - {\bm X}_{1}^{\top} {\bm \beta_{1}} - {\bm U}_{1}^{\top} {\bm \gamma}_{1}^{\star}) \right], \text{if} \; d = 1 \\ 
V_{d} \left[ \frac{Q(\alpha_{d})}{\sigma_{\phi_{d}}^{2}} \sum\limits_{l = 1}^{d-1} A_{dl} {\bm \phi}_{d} + \sum\limits_{l = d + 1}^{D} A_{ld}^{\top} \frac{Q(\alpha_{l})}{\sigma_{\phi_{l}}^{2}} \left( {\bm \phi}_{l} - \sum\limits_{l' = 1, l' \neq d}^{l - 1} A_{ll'} {\bm \phi}_{l'} \right) + \right. \\
\left. \hspace{6cm} + \frac{1}{\sigma_{d}^{2}} ({\bm Y}_{d} - {\bm X}_{d}^{\top} {\bm \beta_{d}} - {\bm U}_{d}^{\top} {\bm \gamma}_{d}^{\star}) \right], \text{if} \; d = 2, \ldots, D - 1 \\ 
V_{d} \left[ \frac{Q(\alpha_{d})}{\sigma_{\phi_{d}}^{2}} \sum\limits_{l = 1}^{d-1} A_{dl} {\bm \phi}_{d} + \frac{1}{\sigma_{d}^{2}} ({\bm Y}_{d} - {\bm X}_{d}^{\top} {\bm \beta_{d}} - {\bm U}_{d}^{\top} {\bm \gamma}_{d}^{\star}) \right], \text{if} \; d = D \end{cases}
\end{align*}
    
    \vspace{0.5cm} \item Update ${\bm \beta}$:
\begin{equation*}
    {\bm \beta} \mid \cdot \sim \text{N} \left(m_{\beta}, V_{\beta} \right),
\end{equation*}
where $V_{\beta} = \left( \sum\limits_{j=1}^{k} \sum\limits_{i:i \in S_{j}} \sum\limits_{t=q}^{T} \frac{X_{it}X_{it}^{\top}}{\sigma_{j}^{2 \star}} + \Sigma_{\beta}^{-1} \right)^{-1}$ and $m_{\beta} = V_{\beta} \left( \sum\limits_{j=1}^{k} \sum\limits_{i:i \in S_{j}} \sum\limits_{t=q}^{T} \frac{( y_{it} - Z_{it}^{\top}{\bm \gamma}_{j}^{\star} - \phi_{i}) X_{it}^{\top}}{\sigma_{j}^{2 \star}} + \Sigma_{\beta}^{-1} \mu_{\beta} \right)$.
\end{enumerate}

\section{Complementary results} \label{appB}

\begin{figure}[H]
 \centerline{\includegraphics[width=\textwidth]{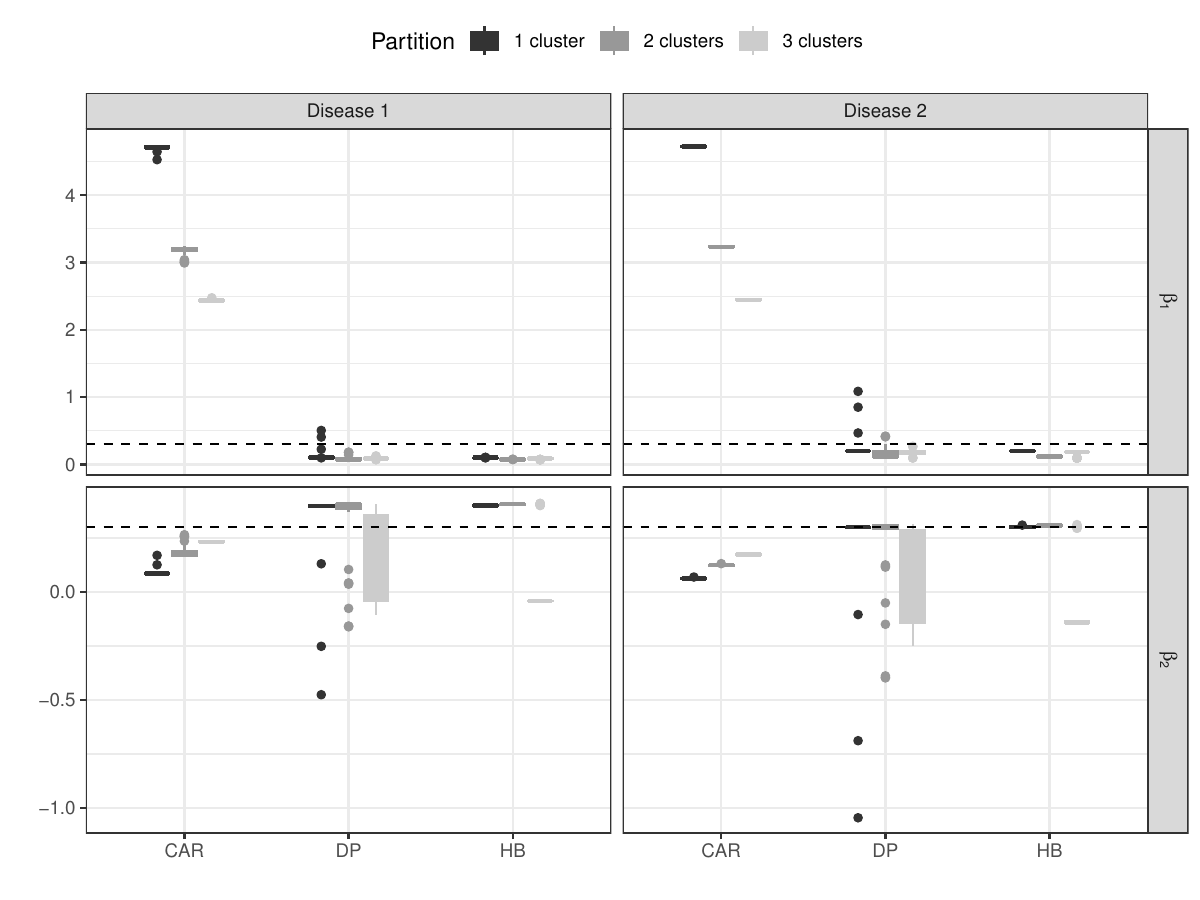}}
\caption{Average of posterior means for the 100 generated datasets. Values obtained by models {\tt MVST.CARar}, PPM-DP, and PPM-HB for scenarios with one, two, and three clusters.}
\label{f:beta_coverage}
\end{figure}

\begin{figure}[H]
 \centerline{\includegraphics[width=\textwidth]{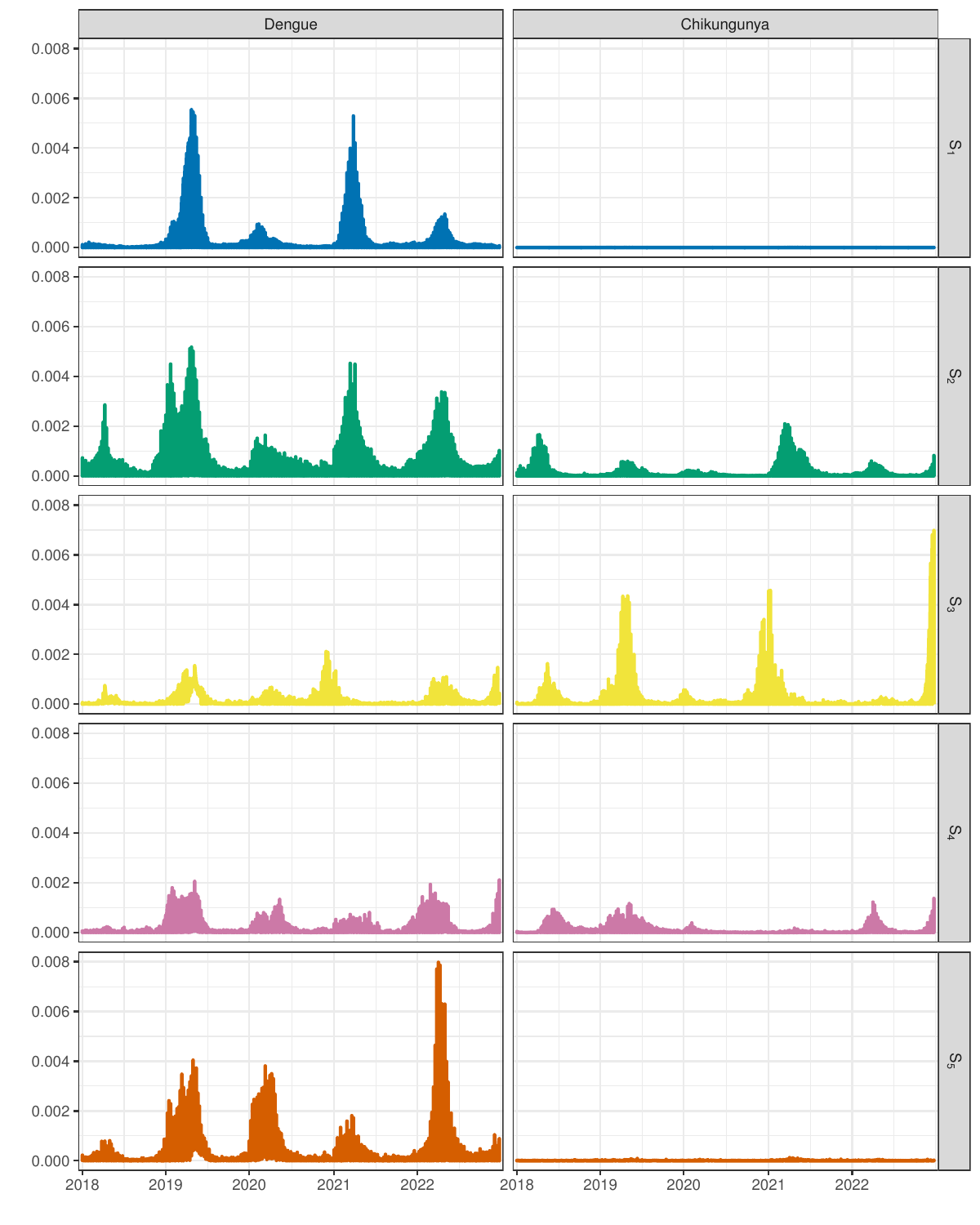}}
\caption{Empirical rate of dengue and chikungunya for areas within each estimated cluster.}
\label{f:empirical_dta}
\end{figure}

\begin{figure}[H]
 \centerline{\includegraphics[width=\textwidth]{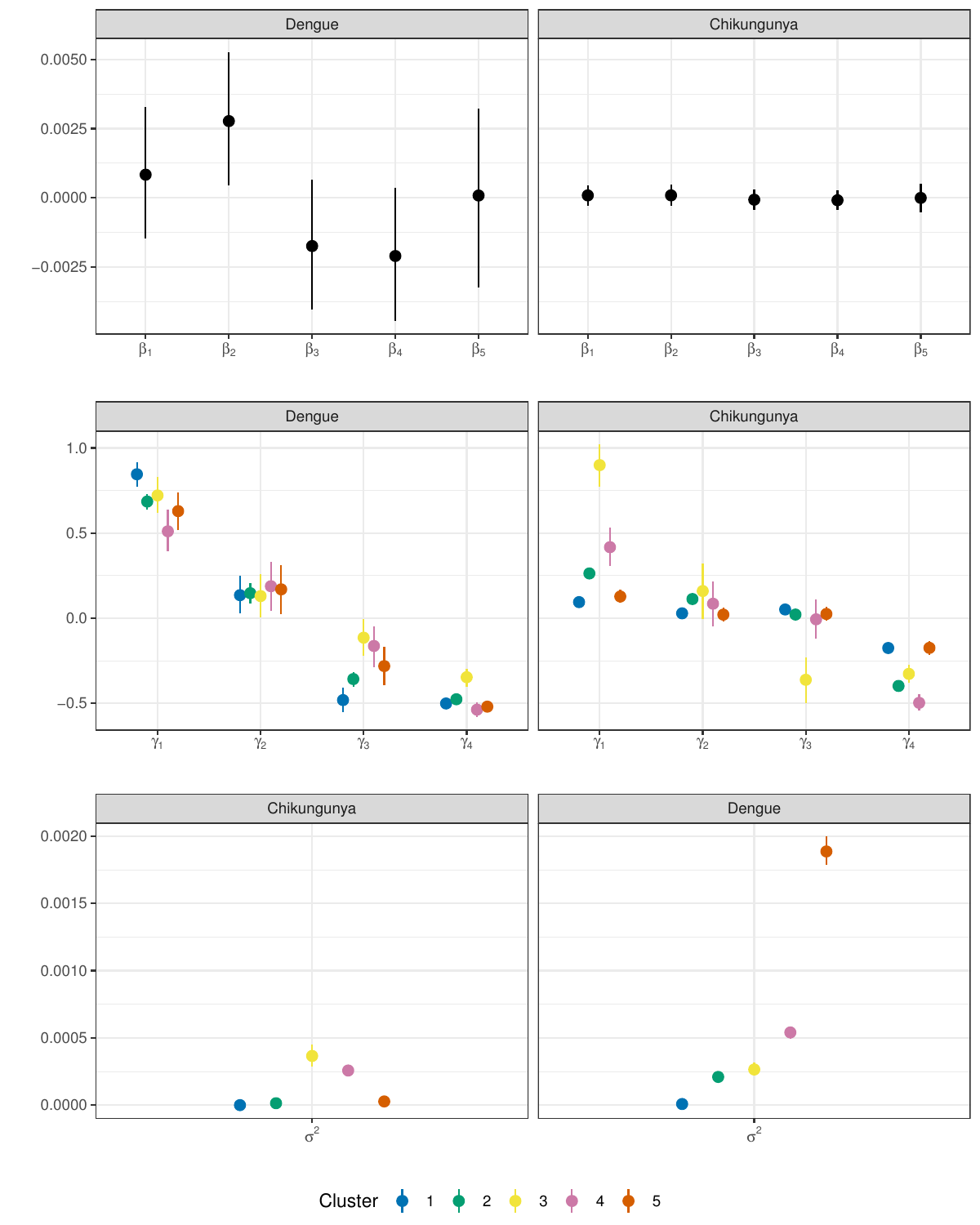}}
\caption{Posterior means and 95\% credible interval of regression coefficients ($\bm \beta$), cluster-specific autoregressive coefficients ($\bm \gamma$), and cluster-specific variances ($\sigma^{2}$). Cluster-specific parameters are conditioned on the estimate partition presented in Figure~\ref{f:est_part}.}
\label{f:cl_parameters}
\end{figure}

Recall from Section~\ref{ss:time_modeling} that autoregressive parameters are modeled as a Normal distribution with mean ${\bm \mu}_{\gamma}$ and covariance matrix ${\bm \Sigma}_{\gamma}$. The posterior mean of these parameters are given by:
\vspace{0.2cm} \begin{equation*}
\hat{\bm \mu}_{\gamma} = \begin{bmatrix}
 0.69 \\ 0.15 \\ -0.27 \\ -0.47 \\ 0.37 \\ 0.08 \\ -0.04 \\ -0.31 \\
\end{bmatrix} \quad \text{and} \quad \hat{\bm \Sigma}_{\gamma} = 
\begin{bmatrix}
 0.74 & 0.49 & 0.62 & 0.62 & 0.62 & 0.61 & 0.62 & 0.62 \\
 0.49 & 0.87 & 0.48 & 0.61 & 0.62 & 0.61 & 0.61 & 0.61 \\
 0.62 & 0.48 & 0.75 & 0.62 & 0.62 & 0.62 & 0.62 & 0.62 \\
 0.62 & 0.61 & 0.62 & 0.62 & 0.62 & 0.61 & 0.62 & 0.62 \\
 0.62 & 0.62 & 0.62 & 0.62 & 0.76 & 0.49 & 0.62 & 0.62 \\
 0.61 & 0.61 & 0.62 & 0.61 & 0.49 & 0.87 & 0.48 & 0.61 \\
 0.62 & 0.61 & 0.62 & 0.62 & 0.62 & 0.48 & 0.74 & 0.61 \\
 0.62 & 0.61 & 0.62 & 0.62 & 0.62 & 0.61 & 0.61 & 0.62 \\
\end{bmatrix}.
\end{equation*} \vspace{0.2cm} 

From the posterior distribution of ${\bm \Sigma}_{\gamma}$, we can obtain the correlation matrix, whose posterior mean is given by:
\vspace{0.2cm} \begin{equation*}
\begin{bmatrix}
1.00 &  & & & & & & \\
-0.07 & 1.00 & & & & & & \\
0.59 & -0.08 & 1.00 & & & & & \\
0.63 & 0.42 & 0.63 & 1.00 & & & & \\
0.45 & 0.30 & 0.44 & 0.62 & 1.00 & & & \\
0.30 & 0.22 & 0.32 & 0.43 & -0.07 & 1.00 & & \\
0.46 & 0.30 & 0.44 & 0.63 & 0.56 & -0.08 & 1.00 & \\
0.64 & 0.41 & 0.64 & 0.98 & 0.62 & 0.44 & 0.62 & 1.00 \\
\end{bmatrix}.
\end{equation*}

\newpage
\section{Univariate analysis} \label{appC}

The model described in Section~\ref{s:model} can be easily adapted to analyze each disease independently. Regression coefficients vector, cluster-specific autoregressive coefficients vector, and variance parameter only suffer a reduction in their dimensions, so that ${\bm \beta}= \{ \beta_{1}, \ldots, \beta_{p} \}$, ${\bm \gamma}_{j}^{\star} = \{ \gamma_{j1}^{\star}, \ldots, \gamma_{jq}^{\star} \}$, and $\sigma_{j}^{2 \star}$ is now a scalar. Thus, prior distributions for these parameters remain the same as presented in \eqref{eq:priors} and \eqref{eq:gamma_prior}. The major model change stands in the specification of the spatial effect vector that is now modeled as an univariate DAGAR, i.e., Gaussian distribution with zero mean and covariance matrix $\sigma^{2}_{\phi} {\bm Q}^{-1}(\alpha)$, where ${\bm Q}(\alpha) = ({\bm I} - {\bm B})^{\top} {\bm \Lambda} ({\bm I} - {\bm B})$. Finally, the model reformulated to accommodate univariate responses is given by:
\begin{align*}  
(y_{it} \mid X_{it}, {\bm \beta}, Z_{it}, {\bm \gamma}_{c_{i}}^{\star}, \phi_{i}, \sigma_{c_{i}}^{2 \star}) & \overset{iid}{\sim} \text{N}(X_{it}^{\top}{\bm \beta} + Z_{it}^{\top}{\bm \gamma}_{c_{i}}^{\star} + \phi_{i}, \, \sigma_{c_{i}}^{2 \star}) \\
{\bm \beta} & \sim \text{N}_{p} ({\bm \mu}_{\beta}, {\bm \Sigma}_{\beta}) \\
(\sigma_{j}^{2 \star} \mid \xi) & \overset{iid}{\sim} \text{inv-Gamma} \left( \nu, \nu \xi \right) \\
\xi & \sim \text{Gamma}(a_{\xi}, b_{\xi}) \\
C(S_{j}) & = \eta^{\ell(S_{j})} \\
({\bm \gamma}_{j}^{\star} \mid {\bm \mu}_{\gamma}, {\bm \Sigma}_{\gamma} ) & \overset{iid}{\sim} \text{N}_{q} \left({\bm \mu}_{\gamma}, {\bm \Sigma}_{\gamma} \right) \\
{\bm \mu}_{\gamma} & \sim \text{N}_{q} ({\bm \mu}_{\mu}, {\bm \Sigma}_{\mu}) \\
{\bm \Sigma}_{\gamma} & \sim \text{inv-Wishart}(\text{df}, {\bm S}) \\ 
({\bm \phi} \mid \; {\bm \alpha}, {\bm \sigma}^{2}_{\phi}) & \sim \text{N}_{n} \Big({\bm \phi}_{1}; {\bm 0}, \sigma^{2}_{\phi} {\bm Q}^{-1}(\alpha) \Big) \\
\alpha & \overset{iid}{\sim} \text{Beta}(a_{\alpha}, b_{\alpha}) \\
\sigma^{2}_{\phi} & \overset{iid}{\sim} \text{inv-Gamma}(a_{\phi}, b_{\phi}),  
 \end{align*}
where $y_{it}$ is now the univariate continuous outcome for area $i$ at time $t$. Worth highlighting that the prior distribution for the random partition remains same as in \eqref{eq:cohesionHB}. This happens because the likelihood plays an important role in partition sampling, therefore, it is in the likelihood that the change lies. The rest of prior distributions remain as defined before. Thus, the posterior distribution is given by:
\begin{align*}
    \nonumber p({\bm \Omega} \mid \; & {\bm Y}, {\bm X}) \propto \left[\prod_{j = 1}^{k} \prod_{i:i \in S_{j}} \prod_{t = q}^{T} \text{N} \left(y_{it}; X_{it}^{\top}{\bm \beta} + Z_{it}^{\top}{\bm \gamma}_{j}^{\star} + \phi_{i}, \sigma_{j}^{\star} \right) \right] \text{N}_{p}\Big({\bm \beta}; {\bm \mu}_{\beta}, {\bm \Sigma}_{\beta} \Big) \\ 
   \nonumber & \times \left[ \prod_{j=1}^{k} \text{N}_{q} \Big({\bm \gamma}_{j}^{\star}; {\bm \mu}_{\gamma}, {\bm \Sigma}_{\gamma} \Big) \right] \text{N}_{q} \Big({\bm \mu}_{\gamma}; {\bm \mu}_{\mu},  {\bm \Sigma}_{\mu} \Big) \; \text{inv-Wishart} \Big({\bm \Sigma}_{\gamma}; \text{df}, {\bm S} \Big) \\ 
    \nonumber & \times \text{N}_{n} \Big({\bm \phi}; {\bm 0}, \sigma^{2}_{\phi} {\bm Q}^{-1}(\alpha) \Big) \; \text{Beta}(\alpha; a_{\alpha}, b_{\alpha}) \; \text{inv-Gamma} (\sigma^{2}_{\phi}; a_{\phi}, b_{\phi} ) \\
    & \times \left[ \prod_{j=1}^{k} \text{inv-Gamma} \Big(\sigma_{j}^{2 \star}; \nu, \nu \xi \Big) \right] \text{Gamma} \Big( \xi; a_{_{\xi}}, b_{_{\xi}} \Big).
\end{align*} 

A simplified version of the algorithm presented in Appendix~\ref{appA} is used to perform posterior simulation via MCMC, where: $\omega$'s update is disregarded; $\alpha$'s and $\sigma^{2}_{\phi}$'s updates consider only the case for $d=1$; and ${\bm \phi}$ is updated as follows:
\begin{equation*}
    {\bm \phi} \mid \cdot \sim \text{N} \Big( m, V \Big),
\end{equation*}
where
\begin{equation*}
V = \left( \frac{1}{\sigma_{\phi}^{2}} Q(\alpha) + \frac{1}{\sigma^{2}} I_{n} \right)^{-1} \quad \text{and} \quad m  = V \left[ \frac{1}{\sigma_{\phi}^{2}} Q(\alpha) + \frac{1}{\sigma^{2}} ({\bm Y} - {\bm X}^{\top} {\bm \beta} - {\bm Z}^{\top} {\bm \gamma}^{\star}) \right].
\end{equation*}

\begin{figure}[H]
 \centerline{\includegraphics[width=3.2in]{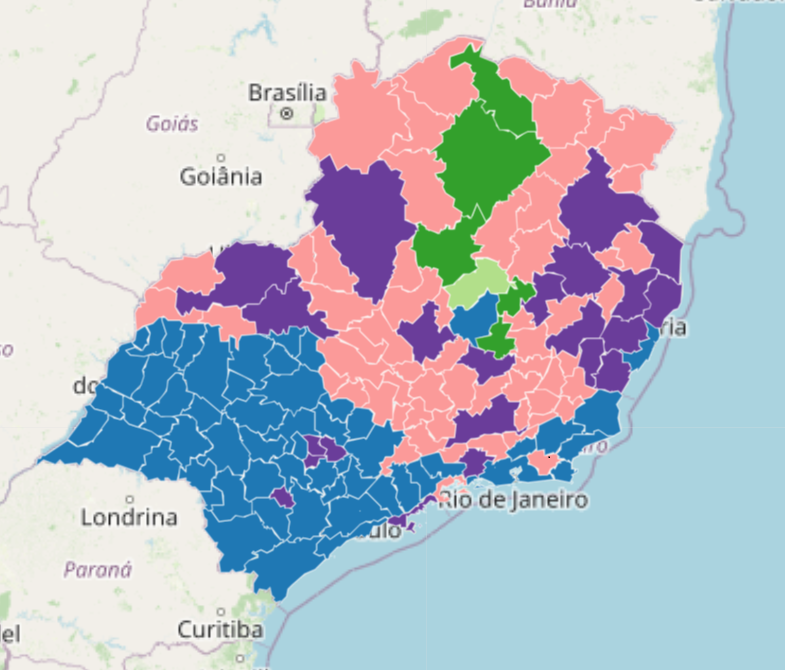}
 \includegraphics[width=3.2in]{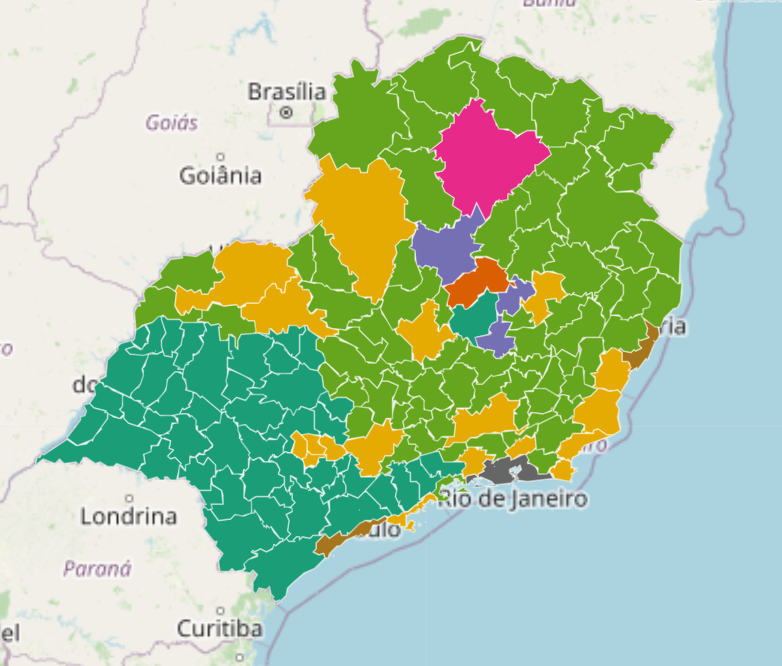}}
\caption{Posterior estimate of the random partition to the Brazilian Southeast region obtained by minimizing the variation of information loss function with identical cost parameters for misclassification. Dengue on the left and chikungunya on the right.}
\label{f:est_part_app}
\end{figure}

\begin{figure}[H]
 \centerline{\includegraphics[width=\textwidth]{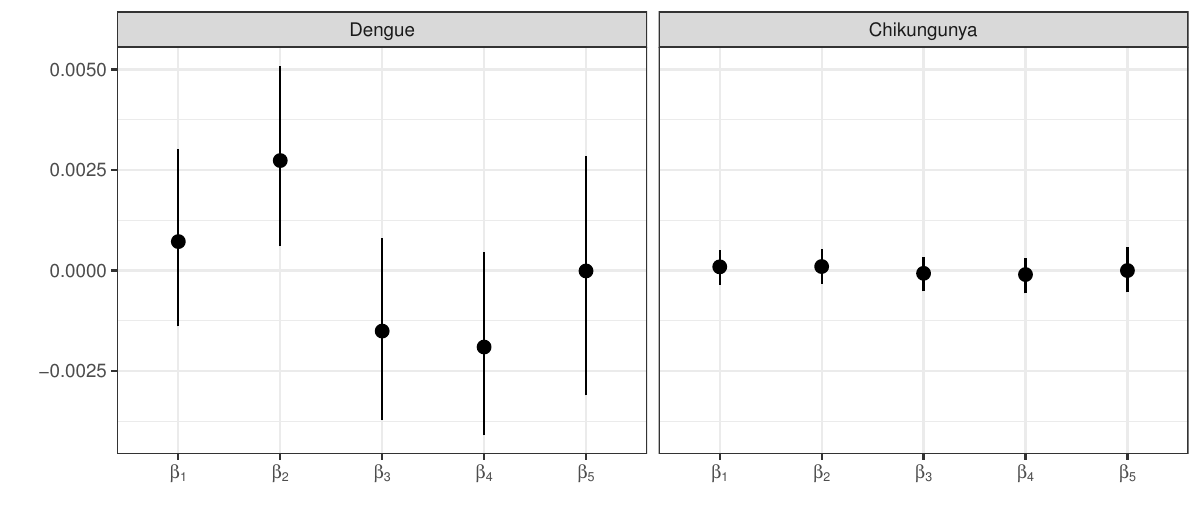}}
\caption{Posterior means and 95\% credible interval of regression coefficients ($\bm \beta$) for dengue and chikungunya.}
\label{f:betas}
\end{figure}

Estimate partition for dengue and chikungunya are displayed in Figure~\ref{f:est_part_app} while posterior means and 95\% credible interval of regression coefficients for both diseases are presented in Figure~\ref{f:betas}.

\end{document}